\documentclass[sigconf]{acmart}
\usepackage[printonlyused,nolist]{acronym}
\usepackage[capitalise,nameinlink]{cleveref}
\usepackage[group-separator={,},group-minimum-digits=4]{siunitx}
\usepackage{orcidlink}

\makeatletter
\renewcommand*\AC@acs[1]{%
    \expandafter\AC@get\csname fn@#1\endcsname\@firstoftwo{#1}}
\makeatother

\usepackage{multirow}
\usepackage{graphicx}

\usepackage[false]{anonymous-acm}

\usepackage{tikz}
\usepackage[]{pgfplots}
\usepackage{pgfplotstable}
\usepackage{fontawesome5}
\usetikzlibrary[
  patterns,
  positioning,
  fit,
  matrix,
  arrows.meta,
  calligraphy,
  math
]
\tikzmath{
  function symlog(\x,\a,\b){
    real \yLarge, \ySmall;
    \yLarge = ((\x>\a) - (\x<-\a)) * (ln(max(abs(\x/\a),1)) + \b);
    \ySmall = (\x >= -\a) * (\x <= \a) * (\x / \a) * \b;
    return \yLarge + \ySmall;
  };
  function symexp(\y,\a,\b){
    real \xLarge, \xSmall;
    \xLarge = ((\y>\b) - (\y<-\b)) * \a * exp(abs(\y) - \b);
    \xSmall = (\y >= -\b) * (\y <= \b) * \a * (\y / \b);
    return \xLarge + \xSmall;
  };
}

\pgfplotsset{compat=1.18}
\pgfplotsset{
    y axis style/.style={
        yticklabel style=#1,
        ylabel style=#1,
        y axis line style=#1,
        ytick style=#1
    },
    General/.style={
        font=\footnotesize,
        width=\linewidth,
        height=3.4cm,
        xtick pos=left,
        xtick align=outside,
        ytick pos=left,
        ytick align=outside,
        ymajorgrids=true,
        grid style=dashed,
        legend cell align={left},
        style=thin,
    },
    BarConfig/.style={
        General,
        ymin=0,
        enlarge x limits=0.25,
        xtick=data,
        xticklabel style={rotate=30, anchor=east},
        xlabel shift=-6pt,
        bar width=8pt,
    },
    ScatterConfig/.style={
      General,
      only marks,
      xtick=data,
      xticklabel style={rotate=90, anchor=east},
    }
}

\definecolor{rwth}   {RGB}{  0  84 159}
\definecolor{rwth-75}{RGB}{ 64 127 183}
\definecolor{rwth-50}{RGB}{142 186 229}
\definecolor{rwth-25}{RGB}{199 221 242}
\definecolor{rwth-10}{RGB}{232 241 250}

\definecolor{black}   {RGB}{  0   0   0}
\definecolor{black-75}{RGB}{100 101 103}
\definecolor{black-50}{RGB}{156 158 159}
\definecolor{black-25}{RGB}{207 209 210}
\definecolor{black-10}{RGB}{236 237 237}

\definecolor{magenta}   {RGB}{227   0 102}
\definecolor{magenta-75}{RGB}{233  96 136}
\definecolor{magenta-50}{RGB}{241 158 177}
\definecolor{magenta-25}{RGB}{249 210 218}
\definecolor{magenta-10}{RGB}{253 238 240}

\definecolor{yellow}   {RGB}{255 237   0}
\definecolor{yellow-75}{RGB}{255 240  85}
\definecolor{yellow-50}{RGB}{255 245 155}
\definecolor{yellow-25}{RGB}{255 250 209}
\definecolor{yellow-10}{RGB}{255 253 238}

\definecolor{petrol}   {RGB}{  0  97 101}
\definecolor{petrol-75}{RGB}{ 45 127 131}
\definecolor{petrol-50}{RGB}{125 164 167}
\definecolor{petrol-25}{RGB}{191 208 209}
\definecolor{petrol-10}{RGB}{230 236 236}

\definecolor{turkis}   {RGB}{  0 152 161}
\definecolor{turkis-75}{RGB}{  0 177 183}
\definecolor{turkis-50}{RGB}{137 204 207}
\definecolor{turkis-25}{RGB}{202 231 231}
\definecolor{turkis-10}{RGB}{235 246 246}

\definecolor{grun}   {RGB}{ 87 171  39}
\definecolor{grun-75}{RGB}{141 192  96}
\definecolor{grun-50}{RGB}{184 214 152}
\definecolor{grun-25}{RGB}{221 235 206}
\definecolor{grun-10}{RGB}{242 247 236}

\definecolor{maigrun}   {RGB}{189 205   0}
\definecolor{maigrun-75}{RGB}{208 217  92}
\definecolor{maigrun-50}{RGB}{224 230 154}
\definecolor{maigrun-25}{RGB}{240 243 208}
\definecolor{maigrun-10}{RGB}{249 250 237}

\definecolor{orange}   {RGB}{246 168   0}
\definecolor{orange-75}{RGB}{250 190  80}
\definecolor{orange-50}{RGB}{253 212 143}
\definecolor{orange-25}{RGB}{254 234 201}
\definecolor{orange-10}{RGB}{255 247 234}

\definecolor{rot}   {RGB}{204   7  30}
\definecolor{rot-75}{RGB}{216  92  65}
\definecolor{rot-50}{RGB}{230 150 121}
\definecolor{rot-25}{RGB}{243 205 187}
\definecolor{rot-10}{RGB}{250 235 227}

\definecolor{bordeaux}   {RGB}{161  16  53}
\definecolor{bordeaux-75}{RGB}{182  82  86}
\definecolor{bordeaux-50}{RGB}{205 139 135}
\definecolor{bordeaux-25}{RGB}{229 197 192}
\definecolor{bordeaux-10}{RGB}{245 232 229}

\definecolor{lila}   {RGB}{122 111 172}
\definecolor{lila-75}{RGB}{155 145 193}
\definecolor{lila-50}{RGB}{188 181 215}
\definecolor{lila-25}{RGB}{222 218 235}
\definecolor{lila-10}{RGB}{242 240 247}

\definecolor{violett}   {RGB}{ 97  33  88}
\definecolor{violett-75}{RGB}{131  78 117}
\definecolor{violett-50}{RGB}{168 133 158}
\definecolor{violett-25}{RGB}{210 192 205}
\definecolor{violett-10}{RGB}{237 229 234}

\usepackage{array}
\usepackage{arydshln}
\usepackage{soul}
\soulregister\acrlong7
\soulregister\cite7

\usepackage{caption}
\usepackage{subcaption}
\usepackage{xspace}
\usepackage{listings}

\lstdefinestyle{rwth-style}{
    basicstyle=\ttfamily\footnotesize,
    breakatwhitespace=true,
    breaklines=true,
    captionpos=b,
    commentstyle=\color{grun},
    escapechar=\%,
    floatplacement=htb,
    frame=none,
    keepspaces=true,
    keywordstyle=\color{rwth},
    language=C++,
    numbers=left,
    numbersep=8pt,
    numberstyle=\tiny\color{black-75},
    numberblanklines=false,
    showspaces=false,
    showstringspaces=false,
    showtabs=false,
    stringstyle=\color{bordeaux},
    tabsize=2,
    xleftmargin=2em,
}
\crefname{lstlisting}{Code}{Codes}
\Crefname{lstlisting}{Code}{Codes}

\AtBeginDocument{%
  \providecommand\BibTeX{{%
    \normalfont B\kern-0.5em{\scshape i\kern-0.25em b}\kern-0.8em\TeX}}}

\copyrightyear{2024} 
\acmYear{2024} 
\setcopyright{acmlicensed}\acmConference[RAPIDO '24]{Rapid Simulation and Performance Evaluation for Design}{January 18, 2024}{Munich, Germany}
\acmBooktitle{Rapid Simulation and Performance Evaluation for Design (RAPIDO '24), January 18, 2024, Munich, Germany}
\acmDOI{10.1145/3642921.3642924}
\acmISBN{979-8-4007-1791-8/24/01}

\usepackage[pages=some]{background}
\newcommand\copyrightnotice{%
    \backgroundsetup{opacity=1, scale=1, angle=0, contents={
            \color{black}%
            \begin{tikzpicture}[remember picture,overlay]%
                \node[anchor=north,yshift=-10pt,text=gray] at (current page.north) {\large PREPRINT - accepted by the \textit{Rapid Simulation and Performance Evaluation for Design Workshop (RAPIDO '24),}};
            \end{tikzpicture}%
        }%
    }%
    \BgThispage%
}

\begin{document}
\newcommand{\nqcc}{NQC²\xspace}
\newcommand{\uniquepgftag}{tag}
\newcommand{\uniquepgflabel}[1]{\label{\uniquepgftag#1}}
\newcommand{\uniquepgfref}[1]{\ref*{\uniquepgftag#1}}
\newcommand{\signcross}{\textcolor{bordeaux}{\faTimesCircle[regular]}\xspace}
\newcommand{\signtick}{\textcolor{grun}{\faCheckCircle[regular]}\xspace}

\tikzstyle{nodetype} = [rectangle, rounded corners, minimum width=1cm, minimum height=.75cm, text centered, text width=1.1cm, draw=black, fill=white, font={\footnotesize}]
\tikzstyle{label}    = [draw=none, font={\footnotesize}, inner sep=0pt, outer sep=0pt]
\tikzstyle{icon}    = [draw=none, font={\Huge}, inner sep=0pt, outer sep=0pt]

\title{\nqcc: A \underline{N}on-Intrusive \underline{Q}EMU \underline{C}ode \underline{C}overage Plugin}

\authoranon{

  \author{Nils Bosbach}
  \orcid{0000-0002-2284-949X}
  \affiliation{%
    \institution{RWTH Aachen University}
    \city{Aachen}
    \country{Germany}
  }

  \author{Alwalid Salama}
  \orcid{0009-0003-1585-7824}
  \affiliation{%
    \institution{RWTH Aachen University}
    \city{Aachen}
    \country{Germany}
  }

  \author{Lukas J\"unger}
  \orcid{0000-0001-9149-1690}
  \affiliation{%
    \institution{MachineWare GmbH}
    \city{Aachen}
    \country{Germany}
  }

  \author{Mark Burton}
  \orcid{0009-0003-6501-637X}
  \affiliation{%
    \institution{\mbox{Qualcomm Technologies, Inc.}}
    \city{Bordeaux}
    \country{France}
  }

  \author{Niko Zurstraßen}
  \orcid{0000-0003-3434-2271}
  \affiliation{%
    \institution{RWTH Aachen University}
    \city{Aachen}
    \country{Germany}
  }

  \author{Rebecca Pelke}
  \orcid{0000-0001-5156-7072}
  \affiliation{%
    \institution{RWTH Aachen University}
    \city{Aachen}
    \country{Germany}
  }

  \author{Rainer Leupers}
  \orcid{0000-0002-6735-3033}
  \affiliation{%
    \institution{RWTH Aachen University}
    \city{Aachen}
    \country{Germany}
  }
}

\renewcommand{\shortauthors}{\textanon{Bosbach, et al.}{Anonymous Author(s)}}
\newcommand*\circled[1]{\tikz[baseline=(char.base)]{
    \node[shape=circle,draw,fill=white,inner sep=1pt] (char) {#1};}}

\begin{abstract}
  \acused{qemu}
  Code coverage analysis has become a standard approach in software development, facilitating the assessment of test suite effectiveness, the identification of under-tested code segments, and the discovery of performance bottlenecks.
  When code coverage of software for embedded systems needs to be measured, conventional approaches quickly meet their limits.
  A commonly used approach involves instrumenting the source files with added code that collects and dumps coverage information during runtime.
  This inserted code usually relies on the existence of an operating and a file system to dump the collected data.
  These features are not available for bare-metal programs that are executed on embedded systems.

  To overcome this issue, we present \nqcc, a plugin for \ac{qemu}.
  \nqcc extracts coverage information from \ac{qemu} during runtime and stores them into a file on the host machine.
  This approach is even compatible with modified \ac{qemu} versions and does not require target-software instrumentation.
  \nqcc outperforms a comparable approach from Xilinx by up to \SI{8.5}{x}.
  \acresetall
\end{abstract}





\maketitle
\copyrightnotice

\begin{acronym}[AAPCS64]
    \acro{aapcs64}[AAPCS64]{Procedure Call Standard for the \acl{aarch64}}
    \acro{aarch64}[AArch64]{\acs{arm} 64-bit Architecture}
    \acro{ai}[AI]{Artificial Intelligence}
    \acro{aiba}[AIBA]{An Automated Intra-cycle Behavioral Analysis for SystemC-based design exploration}
    \acro{algol60}[ALGOL~60]{Algorithmic Language 1960}
    \acro{amd}[AMD]{Advanced Micro Devices}
    \acro{api}[API]{Application Programming Interface}
    \acro{arm}[ARM]{Advanced \acs{risc} Machines}
    \acro{ast}[AST]{Abstract Syntax Tree}
    \acro{avp64}[AVP64]{\acs{arm}v8 \acl{vp}}
    \acro{bb}[BB]{Basic Block}
    \acro{bl}[BL]{Branch-With-Link}
    \acro{bp}[BP]{breakpoint}
    \acro{clint}[CLINT]{Core-local Interrupt Controller}
    \acro{cpu}[CPU]{Central Processing Unit}
    \acro{crc}[CRC]{Cyclic Redundancy Check}
    \acro{csv}[CSV]{Character-Separated Values}
    \acro{db}[DB]{database}
    \acro{dbt}[DBT]{Dynamic Binary Translation}
    \acro{dbms}[DBMS]{\Acl{db} Management System}
    \acro{ddr}[DDR]{Double Data Rate}
    \acro{des}[DES]{Discrete Event Simulation}
    \acro{dla}[DLA]{Deep Learning Accelerator}
    \acro{dmi}[DMI]{Direct Memory Interface}
    \acro{ds}[DS]{Developer Studio}
    \acro{dwarf}[DWARF]{Debugging With Arbitrary Record Formats}
    \acro{eda}[EDA]{Electronic Design Automation}
    \acro{eembc}[EEMBC]{Embedded Microprocessor Benchmark Consortium}
    \acro{etrace}[etrace]{Execution Trace}
    \acro{elf}[ELF]{Executable and Linkable Format}
    \acro{elog}[elog]{Execution Log}
    \acro{esa}[ESA]{European Space Agency}
    \acro{esl}[ESL]{Electronic System Level}
    \acro{fd}[fd]{file descriptor}
    \acro{fig}[FIG]{Fault Injection in glibc}
    \acro{fp}[FP]{Frame Pointer}
    \acro{fpga}[FPGA]{Field Programmable Gate Array}
    \acro{fss}[FSS]{Full-System Simulator}
    \acro{fvp}[FVP]{Fixed Virtual Platform}
    \acro{gcc}[GCC]{GNU Compiler Collection}
    \acro{gdb}[GDB]{GNU Debugger}
    \acro{gic}[GIC]{Generic Interrupt Controller}
    \acro{got}[GOT]{Global Offset Table}
    \acro{gpu}[GPU]{Graphics Processing Unit}
    \acro{gui}[GUI]{Graphical User Interface}
    \acro{hart}[hart]{Hardware Thread}
    \acro{html}[HTML]{HyperText Markup Language}
    \acro{hw}[HW]{Hardware}
    \acro{ibm}[IBM]{International Business Machines}
    \acro{id}[ID]{identifier}
    \acro{ieee}[IEEE]{Institute of Electrical and Electronics Engineers}
    \acro{io}[I/O]{Input/Output}
    \acro{ir}[IR]{Intermediate Representation}
    \acro{irq}[IRQ]{Interrupt Request}
    \acro{isa}[ISA]{Instruction-Set Architecture}
    \acro{iss}[ISS]{Instruction-Set Simulator}
    \acro{lr}[LR]{Link Register}
    \acro{lt}[LT]{Loosely-Timed}
    \acro{mips}[MIPS]{Million Instructions Per Second}
    \acro{miso}[MISO]{Master Input Slave Output}
    \acro{mnist}[MNIST]{Modified National Institute of Standards and Technology}
    \acro{mosi}[MOSI]{Master Output Slave Input}
    \acro{nas}[NAS]{\acs{nasa} Advanced Supercomputing}
    \acro{nasa}[NASA]{National Space Agency}
    \acro{nistt}[NISTT]{A Non-Intrusive SystemC-TLM 2.0 Tracing Tool}
    \acro{npb}[NPB]{\acs{nas} Parallel Benchmarks}
    \acro{nvdla}[NVDLA]{NVIDIA \acl*{dla}}
    \acro{openmp}[OpenMP]{Open Multi-Processing}
    \acro{os}[OS]{Operating System}
    \acrodefplural{os}[OS's]{Operating Systems}
    \acro{pc}[PC]{Program Counter}
    \acro{pccts}[PCCTS]{Purdue Compiler Construction Tool Set}
    \acro{pci}[PCI]{Peripheral Component Interconnect}
    \acro{pid}[PID]{process ID}
    \acro{plic}[PLIC]{Platform-Level Interrupt Controller}
    \acro{plt}[PLT]{Procedure Linkage Table}
    \acro{pmu}[PMU]{Performance Monitoring Unit}
    \acro{pthread}[p\-thread]{\acs{posix} Thread}
    \acro{posix}[POSIX]{Portable Operating System Interface}
    \acro{qemu}[QEMU]{Quick Emulator}
    \acro{ram}[RAM]{Random-Access Memory}
    \acro{risc}[RISC]{Reduced Instruction Set Computer}
    \acro{rtf}[RTF]{Real-Time Factor}
    \acro{rtl}[RTL]{Register-Transfer Level}
    \acro{rtos}[RTOS]{real-time operating system}
    \acro{sata}[SATA]{Serial AT Attachment}
    \acro{sd}[SD]{Secure Digital}
    \acro{sdhci}[SDHCI]{\acs{sd} Host Controller Interface}
    \acro{simd}[SIMD]{Single Instruction, Multiple Data}
    \acro{soc}[SoC]{System-on-a-Chip}
    \acrodefplural{soc}[SoCs]{Systems-on-a-Chip}
    \acro{sp}[SP]{Stack Pointer}
    \acro{spi}[SPI]{Serial Peripheral Interface}
    \acro{sql}[SQL]{Structured Query Language}
    \acro{ssd}[SSD]{Solid State Drive}
    \acro{sw}[SW]{Software}
    \acro{tb}[TB]{Translation Block}
    \acro{tcg}[TCG]{Tiny Code Generator}
    \acro{tcp}[TCP]{Transmission Control Protocol}
    \acro{td}[TD]{Temporal Decoupling}
    \acro{tlm}[TLM]{Transaction-Level Modeling}
    \acro{uart}[UART]{Universal Asynchronous Receiver / Transmitter}
    \acro{unix}[UNIX]{Uniplexed Information and Computing Service}
    \acro{vcd}[VCD]{Value Change Dump}
    \acro{vcml}[VCML]{Virtual Components Modeling Library}
    \acro{vcpu}[vCPU]{virtual CPU}
    \acro{viper}[VIPER]{Virtual Platform Explorer}
    \acro{vp}[VP]{Virtual Platform}
    \acro{wfi}[WFI]{Wait For Interrupt}
\end{acronym}

\acused{cpu}
\acused{html}
\acused{qemu}

\section{Introduction}
\label{sec:introduction}
Code coverage analysis is a fundamental practice in software development, serving as a reliable tool for assessing the effectiveness of test suites, identifying under-tested code segments, and pinpointing performance bottlenecks.
It counts how many times each line of a software program is executed during program runtime.
In 1963, Miller and Maloney published this idea of code coverage~\cite{miller_systematic_1963}, which became a standard in industry and research nowadays~\cite{piwowarski_coverage_1993,ivankovic_code_2019}.

In practice, code coverage analysis finds application in several key areas of software development like test coverage analysis~\cite{ivankovic_code_2019}, bottleneck detection, and guidance of fuzzers~\cite{nagy_full-speed_2019}.
While code coverage analysis has a pivotal role in standard software development, its application to the embedded domain causes some challenges.
These challenges arise from the traditional instrumentation-based approach of code coverage analysis, which involves injecting code into the target executable before, during, or after compilation to gather coverage data.
This approach introduces language and compiler dependencies, alters the executable, and, in many cases, necessitates the execution of the target software within an \ac{os} to use system calls for dumping the coverage data.

\begin{figure}[!t]
  \pgfdeclarelayer{bg}
\pgfsetlayers{bg,main}

\begin{tikzpicture}[node distance=0.5cm and 0.4cm]
    \node[nodetype, fill=rwth!25] (qemu) {\acs{qemu}};
    \node[nodetype, below=of qemu, fill=grun!25] (nqcc) {\nqcc};

    \begin{pgfonlayer}{bg}
        \node[nodetype, fill=rwth!10, fit={(qemu) (nqcc)}] (qemu_nqcc) {};
    \end{pgfonlayer}

    \node[icon, right=of qemu] (input) {\faFileCode[regular]};
    \node[label, anchor=north, yshift=-3pt] (input_lbl) at (input.south) {\shortstack{Target\\Software}};

    \node[icon, right=of nqcc] (covout) {\faFile*[regular]};
    \node[label, anchor=north, yshift=-3pt] (covout_lbl) at (covout.south) {\shortstack{\acs{elog}}};

    \node[fit={(qemu) (nqcc) (input) (covout)}] (qemu_nqcc_files) {};

    \node[nodetype, fill=rwth!25, right=of qemu_nqcc_files] (qemu_etrace) {\shortstack{\acs{qemu}-\\\acs{etrace}\\Tool}};
    \node[icon, right=of qemu_etrace] (lcov) {\faFile*[regular]};
    \node[label, anchor=north, yshift=-3pt] (lcov_lbl) at (lcov.south) {\shortstack{lcov\\File}};

    \node[nodetype, fill=rwth!25, right=of lcov] (genhtml) {\shortstack{genHTML\\Tool}};
    \node[icon, right=of genhtml] (htmlrep) {\faFile*[regular]};
    \node[label, anchor=north, yshift=-3pt] (htmlrep_lbl) at (htmlrep.south) {\shortstack{\acs{html}\\Report}};

    \draw[-Latex] (input) -- (qemu);
    \draw[Latex-Latex] (qemu) -- (nqcc);
    \draw[-Latex] (nqcc) -- (covout);
    \draw[-Latex] (input) -- (qemu_etrace);
    \draw[-Latex] (covout) -- (qemu_etrace);
    \draw[-Latex] (qemu_etrace) -- (lcov);
    \draw[-Latex] (lcov) -- (genhtml);
    \draw[-Latex] (genhtml) -- (htmlrep);

\end{tikzpicture}
  \caption{Interaction between \nqcc and \acs{qemu}, involved files, and used postprocessing tools.}
  \label{fig:overview}
\end{figure}
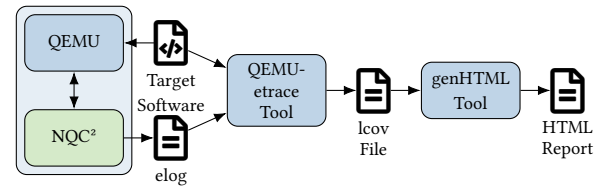

In response to these challenges, this paper introduces an approach that leverages \ac{qemu}~\cite{bellard_qemu_2005}, an open-source simulation software, to enhance code coverage analysis in a non-intrusive and portable manner.
We present the following contributions:

\begin{itemize}
  \item \textbf{\ac{qemu}-\ac{tcg} plugin \nqcc:} We sketch the working principle and implementation details.
  \item \textbf{Performance optimization:} We reduce the slowdown of \nqcc by merging, buffering, and an asynchronous writer.
  \item \textbf{Analysis:} We evaluate the performance of \nqcc for different scenarios showing that we can outperform Xilinx's \ac{qemu}-based coverage solution by a factor of up to \num{8.5}.
\end{itemize}

Our approach addresses the limitations of traditional code coverage analysis and offers a solution that can be easily used with \ac{qemu} implementations that contain custom modifications.
The proposed toolflow is depicted in \cref{fig:overview}.
Our \ac{qemu}-\ac{tcg} plugin \nqcc can be loaded by \ac{qemu} during runtime.
While \ac{qemu} executes the target software, it passes information of the executed code to \nqcc.
The plugin stores the collected data in an \ac{elog} file.
This \ac{elog} file can be processed together with the debugging-symbol-containing \ac{elf} file of the executed target software to generate an \textit{lcov} coverage report.
To visualize this report, the \textit{genHTML}~\cite{linux_test_project_ltp_2023} tool can be used.

\section{Background And Related Work}
\label{sec:background}
Code coverage analysis is nowadays an established tool in industry~\cite{ivankovic_code_2019} that has been used since the late 1960s~\cite{piwowarski_coverage_1993}.
It measures how often a line of a program is executed during runtime.
This information is a valuable insight to the programmer revealing which parts of the software are tested by a testing suite or pinpointing where the performance bottlenecks are.
The collected information can be also used by further approaches like fuzzing~\cite{nagy_full-speed_2019} as guidance.

Coverage data are collected during runtime.
A common approach is the instrumentation of the software before, during or after compilation.
An overview of available code coverage tools is presented in \cref{tab:covtools}.
Modern C/C++ compilers like the \ac{gcc} or \textit{clang} have integrated support for coverage instrumentation during compilation.
They use \textit{gcov}~\cite{free_software_foundation_inc_gcov_2023} and \textit{llvm-cov}~\cite{llvm_project_llvm-cov_2023}, respectly, to instrument the code.
During execution, the instrumentations capture the coverage information and store it in a file.
The produced output can be converted into a coverage report.
Usually, graphical tools are available that create reports that, e.g., display the source code together with the annotation of line-execution counts.
This facilitates the evaluation.
One of those tools is \textit{genHTML} of the \textit{lcov} project~\cite{linux_test_project_ltp_2023}.
It creates a \ac{html}-based report.
Besides C/C++, coverage tools are available for many different languages.

\begin{table}[!b]
  \caption{Code Coverage Approaches.}
  \label{tab:covtools}
  \centering

  \begin{tabular}{l|c|c|c|c}
    \multicolumn{1}{c|}{\multirow{2}{*}{\textbf{Approach}}}                        & \multirow{2}{*}{\textbf{\begin{tabular}[c]{@{}c@{}}Non-\\Intrusive\end{tabular}}} & \multicolumn{2}{c|}{\textbf{Independencies}} & \multirow{2}{*}{\textbf{\begin{tabular}[c]{@{}c@{}}Stand-\\alone\end{tabular}}}              \\
    \multicolumn{1}{c|}{}                                                          &                                                                                   & \textbf{Language}                            & \textbf{OS}                                                                     &            \\ \hline
    gcov~\cite{free_software_foundation_inc_gcov_2023}                             & \signcross                                                                        & \signcross                                   & \signcross                                                                      & \signtick  \\
    llvm-cov~\cite{llvm_project_llvm-cov_2023}                                     & \signcross                                                                        & \signcross                                   & \signcross                                                                      & \signtick  \\
    embedded-gcov~\cite{nasa_jet_propulsion_laboratory_nasa-jplembedded-gcov_2023} & \signcross                                                                        & \signcross                                   & \signtick                                                                       & \signtick  \\
    Blasum et al.~\cite{blasum_gcov_2007}                                          & \signcross                                                                        & \signcross                                   & \signtick                                                                       & \signtick  \\
    Xilinx's \acs{qemu}~\cite{xilinx_xilinxs_2023}                                 & \signtick                                                                         & \signtick                                    & \signtick                                                                       & \signcross \\ \hline
    \nqcc (this work)                                                              & \signtick                                                                         & \signtick                                    & \signtick                                                                       & \signtick
  \end{tabular}
\end{table}

When it comes to coverage analysis for embedded software, additional challenges arise.
Especially for the instrumentation-based approach, standard tools like gcov can only be used if the target software is executed within an \ac{os}.
This is a requirement because the instrumented code depends on system calls such as creating and writing to the file that stores the collected data.
When analyzing the code coverage of bare-metal software, these system calls do not function due to the absence of a file system and \ac{os}.
To circumvent this issue, \textit{embedded-gcov} from the \textit{NASA Jet Propulsion Laboratory}~\cite{nasa_jet_propulsion_laboratory_nasa-jplembedded-gcov_2023} and the approach by Blasum et al.~\cite{blasum_gcov_2007} suggest to directly dump the coverage information into the memory of the embedded target.
After execution, the dump needs to be extracted from the target and stored on a host machine to be analyzed.

The drawbacks of the instrumentation-based approach are that it is programming-language-dependent and changes the target software.
By adding instrumentations to the target software, the binary that is executed when coverage information is collected differs from the one that is executed during normal operation.
This fact can lead to different behavior due to a changed memory layout and added instructions.
The second limitation appears when instrumentations are added by the compiler.
Although it is possible to add instrumentations after compilation~\cite{ben_khadra_efficient_2020}, most state-of-the-art tools like gcov instrument the code before compilation.
This limits the usage of those tools to the specific language they have been developed for.

To overcome this issue, \acp{vp} can be used.
A \ac{vp} is a software-based simulator that mimics the behavior of a full \ac{soc}.
It can be used to develop, run, and analyze the unmodified target software on a host machine like an x86 general-purpose PC.
The instructions of the target program, which have been compiled for the target \ac{isa}, are executed by the \ac{iss}, which is part of the \ac{cpu} model of the \ac{vp}.
Modern \ac{dbt}-based \acp{iss} translate \acp{bb}, which are groups of coherent instructions without branching, from the target \ac{isa} to the host \ac{isa}.
Those translated \acp{bb} are referred to as \acp{tb}.
\acp{tb} can be cached by the simulator so the translation is only needed once per \ac{tb}.
When the same \ac{tb} is executed a second time, the cached version can be used.

A widely-used \ac{vp} that has an \ac{dbt}-based \ac{iss} is \ac{qemu}~\cite{bellard_qemu_2005}.
\ac{qemu} can simulate several target architectures and run on different simulation hosts.
Its internal \ac{iss} is called \ac{tcg}.
\ac{qemu} can be modified to capture tracing data during execution.
This has been done by Xilinx in their \ac{qemu} fork~\cite{xilinx_xilinxs_2023}.
Xilinx added a feature called \ac{etrace} to their \ac{qemu} version, which collects traces from the \ac{tcg} during execution.
These traces are dumped into an \ac{elog} file on the host machine, which can be converted to a lcov file by the \textit{qemu-etrace} tool~\cite{iglesias_edgariglqemu-etrace_2023}.
The strength of this approach is that the target software does not need to be instrumented.
There is no dependency on the compiler or used language and no recompilation is needed for the analysis.
However, the drawback is that Xilinx's solution requires running \ac{qemu} with disabled \ac{tb} chaining.
\ac{tb} chaining is an optimization technique that allows the execution of multiple \acp{tb} without switching back to \ac{qemu}'s main loop in between.
When this feature is disabled, the simulation performance is reduced.

Another drawback is the reliance on Xilinx's customized \ac{qemu} fork.
When the main \ac{qemu} branch receives updates and new features, there can be a significant delay before these changes are incorporated into Xilinx's version.
As of 2023, there are more than \SI{5000}{forks} of the \ac{qemu} GitHub repository~\cite{qemu_qemu_2023}, indicating a strong community interest in enhancing and adapting \ac{qemu} to specific requirements.
Given \ac{etrace}'s deep integration with Xilinx's \ac{qemu}, the ability to reuse this feature in other versions is challenging.

To address the issue of portability, \ac{qemu}'s \ac{tcg} has a plugin feature that has been introduced in version 4.2~\cite{cota_cross-isa_2019}.
A \ac{tcg} plugin is a shared library that can be loaded by \ac{qemu} at runtime.
During this process, \ac{qemu} invokes the \texttt{qemu\_plugin\_install} function of the plugin, which enables the plugin to register callback functions.
A comprehensive overview of the plugin \ac{api} can be found in~\cite{qemu_plugin_api}.
Importantly, plugins can be employed across various \ac{qemu} implementations, ensuring their reusability even when \ac{qemu} has undergone alterations or extensions.
In contrast to intrusive modifications of \ac{qemu}, this feature enables the portability of extensions.

\section{Implementation}
\label{sec:impl}
\nqcc is a \ac{tcg} plugin for \acs{qemu} that leverages the capabilities of the plugin \ac{api} to generate an \acs{elog} file.
This file serves as the foundation for processing by existing tools and facilitates the generation of a comprehensive coverage report as depicted in \cref{fig:overview}.

\begin{figure}[!t]
  \centering
  \tikzstyle{elognode} = [rectangle, minimum width=3cm, minimum height=1em, text width=3cm, draw=black, fill=rwth!10, font={\footnotesize}, anchor=north]
\tikzstyle{eloglbl} = [anchor=west, font={\footnotesize}, xshift=2pt]

\begin{tikzpicture}[node distance=0.3cm and 1cm]
    \node[elognode, fill=rwth!20] (hdr0) {Header: \texttt{etrace\_hdr}};
    \node[elognode, fill=rwth!10] (data0) at (hdr0.south) {Data: \texttt{etrace\_info}};
    \node[elognode, fill=rwth!20] (hdr1) at (data0.south) {Header: \texttt{etrace\_hdr}};
    \node[elognode, fill=rwth!10] (data1) at (hdr1.south) {Data: \texttt{etrace\_arch}};

    \node[elognode, fill=grun!20] (hdr2) at (data1.south) {Header: \texttt{etrace\_hdr}};
    \node[elognode, fill=grun!10] (data2) at (hdr2.south) {Data: \texttt{etrace\_exec}};
    \node[elognode, fill=grun!10] (entry1) at (data2.south) {Entry: \texttt{etrace\_entry64}};
    \node[elognode, fill=grun!10] (entry2) at (entry1.south) {Entry: \texttt{etrace\_entry64}};
    \node[elognode, fill=grun!10] (entry3) at (entry2.south) {...};
    \node[elognode, fill=grun!10] (entry4) at (entry3.south) {Entry: \texttt{etrace\_entry64}};

    \node[elognode, fill=grun!20] (etc1) at (entry4.south) {...};

    \draw [thick, decorate, decoration = {calligraphic brace}] ([yshift=-1pt, xshift=5pt]hdr0.north east) -- node[eloglbl] (block0_lbl) {Version Block} ([yshift=1pt, xshift=5pt]data0.south east);
    \draw [thick, decorate, decoration = {calligraphic brace}] ([yshift=-1pt, xshift=5pt]hdr1.north east) -- node[eloglbl] (block1_lbl) {\shortstack[l]{Architecture\\Block}} ([yshift=1pt, xshift=5pt]data1.south east);
    \draw [thick, decorate, decoration = {calligraphic brace}] ([yshift=-1pt, xshift=1.8cm]hdr0.north east) -- node[eloglbl] (blocks_lbl) {\shortstack[l]{Configuration\\Blocks}} ([yshift=1pt, xshift=1.8cm]data1.south east);
    \draw [thick, decorate, decoration = {calligraphic brace}] ([yshift=-1pt, xshift=5pt]hdr2.north east) -- node[eloglbl] (block2_lbl) {\shortstack[l]{Execution-Data\\Block}} ([yshift=1pt, xshift=5pt]entry4.south east);

\end{tikzpicture}
  \caption{The \acs{elog} file structure.}
  \label{fig:elogschema}
\end{figure}
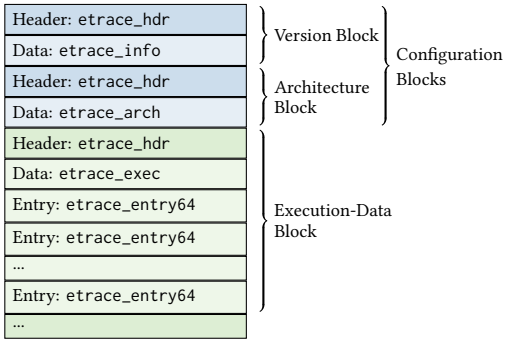

The \ac{elog} file is a binary data file that exhibits a structured layout composed of concatenated blocks as illustrated in \cref{fig:elogschema}.
Each block comprises two main components, a header and a data segment.
The structure of the header, depicted in \cref{lst:hdr}, determines the type and length of the subsequent data segment.
Each data segment type has its own defined structure.

\begin{lstlisting}[float, language=C, caption={\acs{etrace} header struct.}, label=lst:hdr]
struct etrace_hdr {
    uint16_t type;    // type of the subsequent data
    uint16_t unit_id; // CPU ID
    uint32_t len;     // length of the subsequent data
} __attribute__((packed));
\end{lstlisting}

Of particular significance for the coverage evaluation is the entry type presented in \cref{lst:exec}.
For an executed \ac{tb}, the \texttt{etrace\_entry64} struct stores the \texttt{start} and \texttt{end} addresses of the executed instructions within the target's address space and the execution \texttt{duration} in nanoseconds.
To enhance the organization and manageability of the data, multiple \texttt{etrace\_entry64} blocks can be grouped into an execution-data block as visually depicted in \cref{fig:elogschema}.
The \texttt{len} attribute of the \texttt{etrace\_hdr} defines the accumulated sizes of both the \texttt{etrace\_exec} and the \texttt{etrace\_entry64} entries.
During runtime, \nqcc collects the data from \ac{qemu}, stores them in \texttt{etrace\_entry64} structs and dumps them to the \ac{elog} file.

\begin{lstlisting}[float, language=C, caption={Entry for an executed \acs{tb}.}, label=lst:exec]
struct etrace_exec { // type = 1
    uint64_t start_time; // timestamp of first TB exec
} __attribute__((packed));

struct etrace_entry64 {
    uint32_t duration;   // execution duration (ns)
    uint64_t start, end; // start & end addresses
} __attribute__((packed));
\end{lstlisting}

\cref{fig:nqccschema} visualizes how and when \nqcc interacts with \ac{qemu}.
After the plugin has been loaded by \ac{qemu}, the \texttt{qemu\_plugin\_install} function is called.
\nqcc then creates an \ac{elog} file, opens it for writing, and dumps configuration blocks containing version and architecture information.
An asynchronous writer function is executed in a new \ac{pthread} which is from then on used to write data to the \ac{elog} file.
This performance optimization will be discussed in detail in \cref{sec:impl:multi-buffering}.
At the end of the \texttt{qemu\_plugin\_install} function, two callback functions are registered using \ac{qemu}'s \ac{tcg}-plugin \ac{api}.
The first callback function, \texttt{vcpu\_tb\_trans}, notifies \nqcc when the \ac{tcg} translates a new \ac{bb}.
The second callback function, \texttt{at\_exit}, is executed once a \ac{vcpu} exits.

\begin{figure}[!t]
  \centering
  \pgfdeclarelayer{bg}
\pgfsetlayers{bg,main}

\begin{tikzpicture}[node distance=0.3cm and 1cm]
    \node[nodetype, fill=grun!25, text width=3cm, align=left] (init) {\shortstack[l]{
            $\bullet$ Create \& open \acs{elog} file\\
            $\bullet$ Start flusher thread \\
            $\bullet$ Register callbacks
        }};
    \node[nodetype, fill=grun!25, text width=3cm, align=left, below=of init] (tbtrans) {\shortstack[l]{
            $\bullet$ Calc start \& end addresses\\
            $\bullet$ Register callback
        }};
    \node[nodetype, fill=grun!25, text width=3cm, align=left, below=of tbtrans, minimum height=0.5cm] (tbexec) {\shortstack[l]{
            $\bullet$ Store \acs{tb} data in buffer
        }};
    \node[nodetype, fill=grun!25, text width=3cm, align=left, below=of tbexec] (exit) {\shortstack[l]{
            $\bullet$ Flush buffer\\
            $\bullet$ Close \acs{elog} file
        }};
    \node[label, above=of init] (nqcc) {\textbf{\nqcc}};

    \begin{pgfonlayer}{bg}
        \node[nodetype, fill=grun!10, fit={(init) (tbtrans) (tbexec) (exit) (nqcc)}] (nqcc_box) {};
    \end{pgfonlayer}

    \node[label, anchor=base east, xshift=-4.8cm] (qemu) at (nqcc.base west) {\textbf{\acs{qemu}}};
    \node[anchor=south] (qemu_bottom) at (qemu.south|-exit.south) {};

    \begin{pgfonlayer}{bg}
        \node[nodetype, fill=rwth!10, fit={(qemu) (qemu_bottom)}] (qemu_box) {};
    \end{pgfonlayer}

    \draw[Latex-] ([yshift=+1.0em]init.west) -- node[above, anchor=base, yshift=2pt] {\texttt{qemu\_plugin\_install}} ([yshift=+1.0em]qemu_box.east|-init.west);
    \draw[-Latex] ([yshift=+0.0em]init.west) -- node[above, anchor=base, yshift=2pt] {\texttt{vcpu\_tb\_trans\_cb}} ([yshift=+0.0em]qemu_box.east|-init.west);
    \draw[-Latex] ([yshift=-1.0em]init.west) -- node[above, anchor=base, yshift=2pt] {\texttt{at\_exit\_cb}} ([yshift=-1.0em]qemu_box.east|-init.west);

    \draw[Latex-] ([yshift=+0.5em]tbtrans.west) -- node[above, anchor=base, yshift=2pt] {\texttt{vcpu\_tb\_trans}} ([yshift=+0.5em]qemu_box.east|-tbtrans.west);
    \draw[-Latex] ([yshift=-0.5em]tbtrans.west) -- node[above, anchor=base, yshift=2pt] {\texttt{vcpu\_tb\_exec\_cb}} ([yshift=-0.5em]qemu_box.east|-tbtrans.west);

    \draw[Latex-] ([yshift=+0.0em]tbexec.west) -- node[above, anchor=base, yshift=2pt] {\texttt{vcpu\_tb\_exec}} ([yshift=+0.0em]qemu_box.east|-tbexec.west);

    \draw[Latex-] ([yshift=+0.0em]exit.west) -- node[above, anchor=base, yshift=2pt] {\texttt{at\_exit}} ([yshift=+0.0em]qemu_box.east|-exit.west);
\end{tikzpicture}
  \caption{\nqcc schema.}
  \label{fig:nqccschema}
\end{figure}

Every time the \ac{tcg} translates a \ac{bb}, the \texttt{vcpu\_tb\_trans} function is called.
In this function, the addresses of the first and last instructions of the \ac{tb} are calculated and stored in a struct.
A third callback function, \texttt{vcpu\_tb\_exec}, is registered to be called every time the \ac{tb} is executed.
During the registration, a pointer to the struct containing the start and end addresses of the \ac{tb} is handed over.
This pointer is then passed to every call of \texttt{vcpu\_tb\_exec}.

After each execution of a \ac{tb}, the start and end addresses of the \ac{tb} are extracted from the handed-over struct in the \texttt{vcpu\_tb\_exec} function.
The information is copied to an \texttt{etrace\_entry64} struct as shown in \cref{lst:exec}.
The \texttt{etrace\_entry64} struct is placed in a buffer.
Once the buffer is full, the collected \texttt{etrace\_entry64} structs are written to the \ac{elog} file with a preceding header, as depicted in \cref{lst:hdr}, and an \texttt{etrace\_exec} block, as presented in \cref{lst:exec}.

When a \ac{vcpu} exits at the end of the simulation, the remaining \texttt{etrace\_entry64} structs from the buffer are written to the \ac{elog} file.
The file is then closed.
It can be post-processed using the \ac{qemu}-\ac{etrace} tool~\cite{iglesias_edgariglqemu-etrace_2023} to generate a lcov coverage report as shown in \cref{fig:overview}.

The \texttt{etrace\_entry64} structs are collected in a buffer before they are written to the \ac{elog} file.
Buffering the blocks before writing them into the \ac{elog} file reduces the file size by grouping blocks according to \cref{fig:elogschema}.
Thereby, the number of required headers is reduced.
The size of the \ac{elog} file, $S_{elog}$, can be calculated according to \cref{equ:elogfilesize}.

\begin{align}
  S_{elog} & = S_{conf} + \frac{\#TB}{E_{buf}} \cdot \left(S_{hdr} + S_{exec} + E_{buf} \cdot S_{entry64}\right) \label{equ:elogfilesize} \\
           & = \SI{124}{\byte} + \frac{\#TB}{E_{buf}} \cdot \left(\SI{16}{\byte} + E_{buf} \cdot \SI{20}{\byte}\right)                    \\
           & \approx \frac{\#TB}{E_{buf}} \cdot \left(\SI{16}{\byte} + E_{buf} \cdot \SI{20}{\byte}\right) \label{equ:elogfilesizesimple}
\end{align}

The size $S_{elog}$ is composed of a constant amount $S_{conf}$ which is caused by the version and architecture information that is written once at the beginning.
The workload-dependent amount is determined by the number of executed \acp{tb} ($\#TB$), the sizes of the \texttt{etrace\_hdr} ($S_{hdr}$), \texttt{etrace\_exec} ($S_{exec}$), and \texttt{etrace\_entry64} ($S_{entry64}$) structs, and the numer of \texttt{etrace\_entry64} structs that fit into the buffer ($E_{buf}$).
Since the constant part is relatively small and thereby negligible, the equation can be simplified to \cref{equ:elogfilesizesimple}.

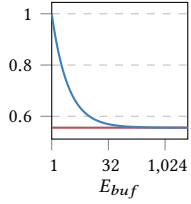
\begin{figure}
  \centering
  \begin{subfigure}[c]{.4\linewidth}
    \begin{tikzpicture}
      \begin{axis}[
          General,
          xmin=1,
          xmax=4096,
          domain=1:4096,
          samples=512,
          xmode=log,
          log basis x=2,
          xticklabel={
              \pgfkeys{/pgf/fpu=true}
              \pgfmathparse{2^(\tick)}%
              \pgfmathprintnumber[fixed relative, precision=5]{\pgfmathresult}
              \pgfkeys{/pgf/fpu=false}
            },
          xlabel={$E_{buf}$},
          xlabel shift=-4pt,
        ]
        \addplot[thick, mark=none, draw=bordeaux-75] {5/9};
        \addplot[thick, mark=none, draw=rwth-75] {4/9 * 1/x + 5/9};
      \end{axis}
    \end{tikzpicture}
  \end{subfigure}\hfill
  \begin{subfigure}[c]{.59\linewidth}
    \begin{equation}
      \frac{S_{elog}(E_{buf})}{S_{elog}(E_{buf} = 1)} \approx \frac{4}{9} \cdot \frac{1}{E_{buf}} + \frac{5}{9} \label{equ:elogfileratio}
    \end{equation}
  \end{subfigure}
  \caption{Reduction of the \ac{elog} file size due to buffering.}
  \label{fig:elogfilesize}
\end{figure}

To estimate the influence of buffering on the \ac{elog} file size, the ratio of the \ac{elog} file size with buffering, $S_{elog}$, to the \ac{elog} file size without buffering, $S_{elog}(E_{buf} = 1)$, can be calculated according to \cref{equ:elogfileratio}.
The resulting relative reduction of the file size is plotted in \cref{fig:elogfilesize}.
It shows that buffering and bundling of the \texttt{etrace\_entry64} structs can reduce the file size of the \ac{elog} file by up to \SI{44}{\percent}.
If only $32$ elements are bundled, the file size is already reduced by \SI{43}{\percent}.
Further bundling has only a limited influence on the file size.

\subsection{Multi-buffering}
\label{sec:impl:multi-buffering}
The buffering and bundling of \texttt{etrace\_entry64} entries serve a dual purpose, benefiting not only in the reduction of the \ac{elog} file size but also in enhancing performance-optimization possibilities.
When the buffer is full, the contained data need to be written to the \ac{elog} file.
While this is done, \ac{qemu} is suspended which reduces the performance.
To circumvent this issue, we suggest the implementation of an asynchronous writer \ac{pthread} together with multiple buffers.

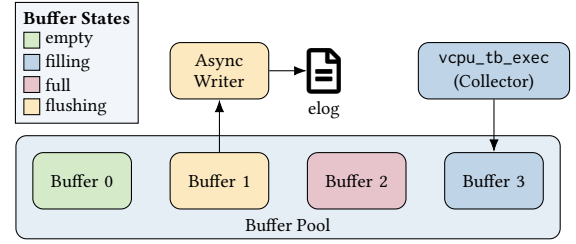
\begin{figure}[!t]
  \pgfdeclarelayer{bg}
\pgfsetlayers{bg,main}

\begin{tikzpicture}[node distance=0.7cm and 0.5cm]
    \node[nodetype, fill=grun!25] (c0) {Buffer 0};
    \node[nodetype, fill=orange!25, right=of c0] (c1) {Buffer 1};
    \node[nodetype, fill=bordeaux!25, right=of c1] (c2) {Buffer 2};
    \node[nodetype, fill=rwth!25, right=of c2] (c3) {Buffer 3};

    \node[fit={(c0) (c1) (c2) (c3)}] (cpinner){};
    \node[label, anchor=north] (cp_lbl) at (cpinner.south) {Buffer Pool};

    \begin{pgfonlayer}{bg}
        \node[nodetype, fill=rwth!10, fit={(cpinner) (cp_lbl)}] (cp) {};
    \end{pgfonlayer}

    \node[nodetype, fill=orange!25, above=of c1] (flusher) {Async Writer};
    \node[icon, right=of flusher] (outfile) {\faFile*[regular]};
    \node[label, anchor=north, yshift=-3pt] (outfile_lbl) at (outfile.south) {\shortstack{\acs{elog}}};
    \node[nodetype, fill=rwth!25, above=of c3, text width=1.8cm] (tbexec) {\shortstack{\texttt{vcpu\_tb\_exec}\\(Collector)}};

    \draw[-Latex] (c1) -- (flusher);
    \draw[-Latex] (flusher) -- (outfile);
    \draw[-Latex] (tbexec) -- (c3);

    \node[label, anchor=north west, xshift=3pt, yshift=12pt] (legend_lbl) at (cp.west|-flusher.north){\textbf{Buffer States}};
    \node[rectangle, draw, fill=grun!25, anchor=north west, yshift=-3pt] (leg_empty) at (legend_lbl.south west) {};
    \node[rectangle, draw, fill=rwth!25, anchor=north west, yshift=-2pt] (leg_filled) at (leg_empty.south west) {};
    \node[rectangle, draw, fill=bordeaux!25, anchor=north west, yshift=-2pt] (leg_full) at (leg_filled.south west) {};
    \node[rectangle, draw, fill=orange!25, anchor=north west, yshift=-2pt] (leg_flushed) at (leg_full.south west) {};

    \node[label, anchor=west, xshift=2pt] (leg_empty_lbl) at (leg_empty.east) {empty};
    \node[label, anchor=west, xshift=2pt] (leg_filled_lbl) at (leg_filled.east) {filling};
    \node[label, anchor=west, xshift=2pt] (leg_full_lbl) at (leg_full.east) {full};
    \node[label, anchor=west, xshift=2pt] (leg_flushed_lbl) at (leg_flushed.east) {flushing};

    \begin{pgfonlayer}{bg}
        \node[draw, fill=rwth!5, fit={(legend_lbl) (leg_empty) (leg_flushed) (leg_full) (leg_filled) (leg_empty_lbl) (leg_flushed_lbl) (leg_full_lbl) (leg_filled_lbl)}] (leged) {};
    \end{pgfonlayer}
\end{tikzpicture}
  \caption{\nqcc multi-buffering schema.}
  \label{fig:multibuf}
\end{figure}

The concept is sketched in \cref{fig:multibuf} for four buffers.
Instead of a single buffer that is filled and flushed once it is full, we have multiple buffers.
A state is assigned to each buffer which can either be \textit{empty}, \textit{filling}, \textit{full}, or \textit{flushing}.
During the initialization of \nqcc, an asynchronous writer \ac{pthread} is spawned (see \cref{fig:nqccschema}).
Each buffer can be accessed from two different \acp{pthread}, the writer or the main \nqcc \ac{pthread}, called \textit{collector} in the following.
At the beginning, all buffers are empty.
The collector changes the state of the first buffer to filling.
It adds \texttt{etrace\_entry64} structs to the buffer until it is full.
Then it changes the state to full.
Once the next buffer is in the empty state, the collector changes the state to filling and continues.
If the next buffer is in the full or flushing state, the collector needs to wait until the state is changed to empty by the writer.

The writer runs in parallel to the collector in the asynchronous \ac{pthread}.
At the beginning of the simulation, it waits until the collector changes the state of the first buffer to full.
Then the writer updates the state to flushing, flushes the data into the \ac{elog} file and sets the state to empty.
It waits until the collector changes the state of the next buffer to full before it continues writing data to the \ac{elog} file.

The waiting of the collector for an empty and the writer for a full buffer is synchronized using POSIX condition variables.
Every time a state is changed from flushing to empty or from filling to full, a condition variable is notified to alert the potentially waiting collector or writer.

\subsection{Merging}
\label{sec:impl:merging}
\nqcc includes the merging of \texttt{etrace\_entry64} structs as a second optimization.
In the \texttt{vcpu\_tb\_exec} callback, before adding a new entry to the buffer, it is checked whether the buffer is empty.
If that is not the case, the \texttt{end} address of the last element in the buffer and the \texttt{start} address of the executed \ac{tb} are compared.
In the case of a match, the last entry in the buffer can be updated instead of adding a new entry to the buffer.
This is done by setting the \texttt{end} address of the last entry to the end address of the current entry.
If timing is annotated, the execution duration of the current \ac{tb} needs to be added to the \texttt{duration} field of the updated entry.

\section{Results}
\label{sec:results}
To measure the slowdown of \nqcc, we assess the code coverage of the widely-used bare-metal benchmarks Coremark~\cite{gal-on_exploring_2012}, Dhrystone~\cite{weicker_dhrystone_1984}, Stream~\cite{mccalpin_memory_1995}, and Whetstone~\cite{wichmann_statistics_1970}.
All benchmarks are evaluated on \ac{qemu} version 8.1.1~\cite{qemu_qemu_2023} with and without \nqcc enabled, and Xilinx's \ac{qemu} fork version 2023.1\_update1~\cite{xilinx_xilinxs_2023}.
The simulated target architecture is \textit{aarch64}.
The used host \ac{cpu} is an AMD Ryzen~9~3900X 12-core processor.
When \nqcc is loaded, the number of buffers, the number of \texttt{etrace\_entry64} structs per buffer, and the merging option can be configured.

\begin{figure}[!t]
  \centering
  \begin{subfigure}[t]{0.45\linewidth}
    \centering
    \mbox{
      \pgfplotstableread[col sep=comma]{plot/data/results_slowdown.csv}\results
\pgfplotstableread[col sep=comma]{plot/data/results_filesize.csv}\resultsfilesize

\begin{tikzpicture}
    \begin{axis}[
            BarConfig,
            ybar,
            axis y line*=left,
            y axis style=rwth,
            yminorgrids=true,
            xticklabels from table={\results}{benchmark},
            ylabel={\acs{qemu} \acs{mips}},
            ylabel shift=-4pt,
            xlabel=Benchmark,
            bar width=6pt,
            bar shift=-3pt,
            ymax=2000,
        ]
        \addplot[fill=rwth-50, postaction={pattern=north east lines, pattern color=white}] table [x expr=\coordindex, y=MIPS_mainline] {\results};
    \end{axis}
    \begin{axis}[
            BarConfig,
            ybar,
            yticklabel pos=right,
            ytick pos=right,
            hide x axis,
            axis y line*=right,
            y axis style=grun,
            yminorgrids=false,
            ymajorgrids=false,
            xticklabels from table={\resultsfilesize}{benchmark},
            ylabel={$\frac{\#\text{translated \acsp{tb}}}{\#\text{executed \acsp{tb}}}$},
            ylabel shift=-4pt,
            bar width=6pt,
            bar shift=+3pt,
            ymax=2e-5,
        ]
        \addplot[fill=grun-50, postaction={pattern=north east lines, pattern color=white}] table [x expr=\coordindex, y=trans_tb_prob] {\resultsfilesize};
    \end{axis}
\end{tikzpicture}
    }
    \caption{\acs{qemu} execution statistics.}
    \label{fig:bench:mips}
  \end{subfigure}\hfill
  \begin{subfigure}[t]{0.45\linewidth}
    \centering
    \mbox{
      \pgfplotstableread[col sep=comma]{plot/data/results_filesize.csv}\results

\begin{tikzpicture}
    \begin{axis}[
            BarConfig,
            ybar,
            yminorgrids=true,
            xticklabels from table={\results}{benchmark},
            ylabel={Proportion},
            ylabel shift=-4pt,
            xlabel=Benchmark,
        ]
        \addplot[fill=rwth-50, postaction={pattern=north east lines, pattern color=white}] table [x expr=\coordindex, y=merge_prob_max] {\results};
    \end{axis}
\end{tikzpicture}
    }
    \caption{Proportion of mergeable \texttt{etrace\_entry64} blocks.}
    \label{fig:bench:merge}
  \end{subfigure}
  \caption{\acs{qemu} execution statistics.}
  \label{fig:bench}
\end{figure}
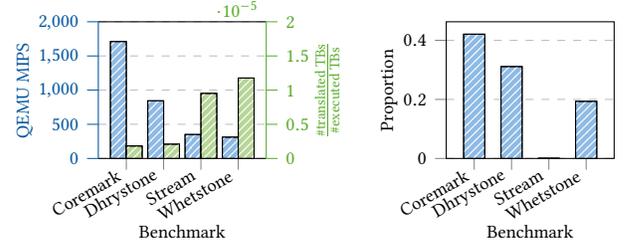

\cref{fig:bench} shows general properties of the benchmarks and the execution by \ac{qemu} that are independent of the code coverage analysis.
\cref{fig:bench:mips} depicts the simulation speed of \ac{qemu} for the different benchmarks measured in \ac{mips}.
It can be seen that the reached simulation speeds differ for the benchmarks.
Coremark and Dhrystone reach the highest simulation speeds.
One reason for the high simulation speed is that both benchmarks are dominated by simple integer-arithmetic-based instructions~\cite{bosbach_entropy-based_2023}.
The higher execution speed for Coremark is caused by the lower amount of load instructions compared to Dhrystone.
Stream mainly uses load and store instructions, which causes a higher slowdown.
Whetstone consists of many floating point operations, which are more complex to simulate than integer operations.

Furthermore, \cref{fig:bench:mips} shows the probability that a \ac{tb} needs to be translated before execution.
Since \ac{qemu} buffers translated \acp{tb}, the probability that a \ac{tb} has been translated in the past and can directly be executed is above \SI{99.9}{\percent} for all tested benchmarks.
However, the \ac{tb}-translation probability of Whetstone is more than \SI{6.4}{x} higher than the one of Coremark.
A lower \ac{tb} probability can lead to a higher performance of \ac{qemu} due to the reduced translations.

\cref{fig:bench:merge} shows how often subsequent \texttt{etrace\_entry64} blocks can be merged (cf. \cref{sec:impl:merging}).
For example, \SI{124257227} of the total \num{295290670} \texttt{etrace\_entry64} blocks (\SI{42.08}{\percent}) that need to be stored for a Coremark execution can be merged with their predecessor.
In contrast, for the Stream benchmark, merging can only be applied to \num{10481} of the \num{55024880} \texttt{etrace\_entry64} blocks (\SI{0.02}{\percent}).

\begin{figure}[!t]
  \centering
  \begin{subfigure}[b]{0.45\linewidth}
    \centering
    \pgfplotstableread[col sep=comma]{plot/data/results_slowdown.csv}\results

\begin{tikzpicture}
    \pgfplotsset{every axis/.style={ 
                BarConfig,
                ybar stacked,
                ymin=0,
                ymax=30,
                minor y tick num=1,
                yminorgrids=true,
                xticklabels from table={\results}{benchmark},
                ylabel={Slowdown $S = \frac{t_{cov}}{t_{QEMU}}$},
                xlabel=Benchmark,
                bar width=6pt,
                transpose legend,
                legend style={at={(1,1.2)}, anchor=south east, legend columns=2, inner sep=1pt, cells={anchor=west}, /tikz/column 2/.style={column sep=5pt}},
            }}
    \begin{axis}[bar shift=-3pt]
        \addplot[fill=rwth-50, postaction={pattern=north east lines, pattern color=white}] table [x expr=\coordindex, y=min_plugin] {\results}; \label{pgf:slowdown:minplugin}
        \addplot[ycomb, xshift=-3pt, mark=-, draw=rwth] table [x expr=\coordindex, y=max_plugin] {\results}; \label{pgf:slowdown:maxplugin}
    \end{axis}
    \begin{axis}[hide axis, bar shift=+3pt]
        \addlegendimage{/pgfplots/refstyle=pgf:slowdown:minplugin}\addlegendentry{\nqcc min}
        \addlegendimage{/pgfplots/refstyle=pgf:slowdown:maxplugin, xshift=3pt}\addlegendentry{\nqcc max}

        \addplot[fill=grun-50, postaction={pattern=north east lines, pattern color=white}] table [x expr=\coordindex, y=max_xilinx] {\results}; \label{pgf:slowdown:xilinx}
        \addlegendimage{/pgfplots/refstyle=pgf:slowdown:maxplugin}\addlegendentry{Xilinx's \acs{etrace}}
    \end{axis}

\end{tikzpicture}
    \caption{Runtime slowdown.}
    \label{fig:res:slowdown}
  \end{subfigure}
  \begin{subfigure}[b]{0.45\linewidth}
    \centering
    \pgfplotstableread[col sep=comma]{plot/data/results_filesize.csv}\results

\begin{tikzpicture}
    \pgfplotsset{every axis/.style={ 
            }}
    \begin{axis}[
            BarConfig,
            ybar stacked,
            ymin=0,
            minor y tick num=1,
            yminorgrids=true,
            xticklabels from table={\results}{benchmark},
            ylabel={Etrace Size (\si{\gibi\byte})},
            xlabel=Benchmark,
            legend style={at={(0.5,1.2)}, anchor=south, legend columns=1, inner sep=1pt},
        ]
        \addplot[fill=rwth-50, postaction={pattern=north east lines, pattern color=white}] table [x expr=\coordindex, y=filesize_min_GiB] {\results};
        \addlegendentry{Merge = on}

        \addplot[fill=rwth-25, postaction={pattern=vertical lines, pattern color=white}] table [x expr=\coordindex, y=filesize_ptp_GiB] {\results};
        \addlegendentry{Merge = off}
    \end{axis}
\end{tikzpicture}
    \caption{Trace file sizes.}
    \label{fig:res:filesize}
  \end{subfigure}
  \caption{\nqcc benchmark results.}
  \label{fig:res}

\end{figure}
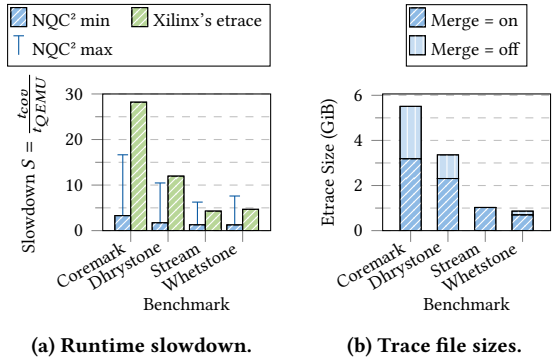

\cref{fig:res:slowdown} shows the runtime slowdowns \nqcc and Xilinx's \ac{qemu} with enabled \ac{etrace} cause compared to \ac{qemu} 8.1.1.
Since the buffer count, buffer size, and merging can be configured for \nqcc and influence the performance, the minimum and maximum values that can be achieved are shown.
The evaluated number of buffers is between \num{1} and \num{16}, the buffers can fit between \num{512} and \num{65536} \texttt{etrace\_entry64} elements, and merging is enabled or disabled.

While Xilinx's \ac{etrace} implementation causes enormous slowdowns by a factor of \num{28.2} for the Coremark benchmark (\cref{fig:res:slowdown}), the slowdown of \nqcc can be reduced to a maximum factor of \num{3.3} for all evaluated benchmarks.
For the Coremark benchmark, \nqcc outperforms Xilinx's solution by \SI{8.5}{x}.
A reason for the better performance of \nqcc is that Xilinx's \ac{etrace} implementation requires disabled \ac{tb} block chaining to work.
Furthermore, \nqcc has, in contrast to Xilinx's implementation, multi-buffering and variable buffer-size capabilities.
Xilinx uses a single buffer that can fit \num{16384} \texttt{etrace\_entry64} elements.
They use a single \ac{pthread}.

The worst-case slowdown of \nqcc exhibits a notable dependency on the executed workload.
When we compare the trends depicted in \cref{fig:bench:mips,fig:res:slowdown}, a consistent pattern emerges in the course of \ac{mips} values and the resulting slowdown.
It seems that \nqcc introduces a higher slowdown when it is used with workloads that can be executed with high \ac{mips} values by \ac{qemu}.
Tuning buffer parameters and merge options can help to limit the slowdown.
For the best-case scenario, \nqcc always outperforms Xilinx's \ac{etrace}.

\cref{fig:res:filesize} shows the file size of the \ac{elog} file for enabled and disabled merging.
According to \cref{equ:elogfileratio}, the file size also depends on the used buffer size.
However, for the evaluated buffer sizes, which are larger than or equal to \SI{512}, the buffer size has a neglectable impact as shown in \cref{fig:elogfilesize}.
Hence, this dependency is not further evaluated.
It can be seen that the \ac{elog} files can rapidly grow to multiple gigabytes.
Merging helps to reduce the file size but the effectiveness depends on the workload.
The impact of merging directly corresponds to the proportion of \texttt{etrace\_entry64} blocks that can be merged as depicted in \cref{fig:bench:merge}.
Benchmarks that have a high probability that blocks can be merged, like Coremark and Dhrystone, can benefit from merging.
Merging cannot reduce the file size for benchmarks with few mergeable blocks, such as Stream.

\begin{figure}
  \renewcommand{\uniquepgftag}{coremark}
  \begin{tikzpicture}
    \matrix [
        draw,
        fill=white,
        matrix of nodes,
        font=\footnotesize,
        anchor=south east,
        row sep=0pt,
        inner sep=0.5pt,
        column 1/.style={anchor=base east},
        line width=0.6pt,
    ] {
        \textsubscript{Merge}\textbackslash\textsuperscript{Buffers} & 1                      & 2                      & 4                      & 8                      & 16                      \\
        on                                                           & \uniquepgfref{pgf:1_1} & \uniquepgfref{pgf:2_1} & \uniquepgfref{pgf:4_1} & \uniquepgfref{pgf:8_1} & \uniquepgfref{pgf:16_1} \\
        off                                                          & \uniquepgfref{pgf:1_0} & \uniquepgfref{pgf:2_0} & \uniquepgfref{pgf:4_0} & \uniquepgfref{pgf:8_0} & \uniquepgfref{pgf:16_0} \\
    };
\end{tikzpicture}
  \begin{subfigure}[b]{0.45\linewidth}
    \centering
    \pgfplotstableread[col sep=comma]{plot/data/results_Coremark_caches_slowdown.csv}\slowdowndata
    \renewcommand{\uniquepgftag}{coremark}
    \mbox{
      \begin{tikzpicture}
    \begin{axis}[
            General,
            xmode=log,
            log basis x=2,
            xticklabel={
                    \pgfkeys{/pgf/fpu=true}
                    \pgfmathparse{2^(\tick)}%
                    \pgfmathprintnumber[fixed relative, precision=5]{\pgfmathresult}
                    \pgfkeys{/pgf/fpu=false}
                },
            xmin=512,
            xmax=65536,
            ymin=0,
            ymax=15,
            xlabel={Buffer size (entries)},
            ylabel={Slowdown $S = \frac{t_{NQC^2}}{t_{QEMU}}$},
            xlabel shift=-4pt,
            ylabel shift=-4pt,
            mark size=1pt,
        ]
        \addplot[thick, dotted, line cap = round, mark=*, mark options={solid}, color=rwth-75] table [x=cache_size, y=slowdown_cache_cnt1_merge0] {\slowdowndata}; \uniquepgflabel{pgf:1_0}
        \addplot[thick, dotted, line cap = round, mark=*, mark options={solid}, color=grun-75] table [x=cache_size, y=slowdown_cache_cnt2_merge0] {\slowdowndata}; \uniquepgflabel{pgf:2_0}
        \addplot[thick, dotted, line cap = round, mark=*, mark options={solid}, color=bordeaux-75] table [x=cache_size, y=slowdown_cache_cnt4_merge0] {\slowdowndata}; \uniquepgflabel{pgf:4_0}
        \addplot[thick, dotted, line cap = round, mark=*, mark options={solid}, color=orange-75] table [x=cache_size, y=slowdown_cache_cnt8_merge0] {\slowdowndata}; \uniquepgflabel{pgf:8_0}
        \addplot[thick, dotted, line cap = round, mark=*, mark options={solid}, color=violett-75] table [x=cache_size, y=slowdown_cache_cnt16_merge0] {\slowdowndata}; \uniquepgflabel{pgf:16_0}

        \addplot[thick, mark=*, color=rwth-75] table [x=cache_size, y=slowdown_cache_cnt1_merge1] {\slowdowndata}; \uniquepgflabel{pgf:1_1}
        \addplot[thick, mark=*, color=grun-75] table [x=cache_size, y=slowdown_cache_cnt2_merge1] {\slowdowndata}; \uniquepgflabel{pgf:2_1}
        \addplot[thick, mark=*, color=bordeaux-75] table [x=cache_size, y=slowdown_cache_cnt4_merge1] {\slowdowndata}; \uniquepgflabel{pgf:4_1}
        \addplot[thick, mark=*, color=orange-75] table [x=cache_size, y=slowdown_cache_cnt8_merge1] {\slowdowndata}; \uniquepgflabel{pgf:8_1}
        \addplot[thick, mark=*, color=violett-75] table [x=cache_size, y=slowdown_cache_cnt16_merge1] {\slowdowndata}; \uniquepgflabel{pgf:16_1}
    \end{axis}
\end{tikzpicture}
    }
    \caption{Coremark.}
    \label{fig:slowdown:coremark}
  \end{subfigure}
  \begin{subfigure}[b]{0.45\linewidth}
    \centering
    \pgfplotstableread[col sep=comma]{plot/data/results_Dhrystone_caches_slowdown.csv}\slowdowndata
    \renewcommand{\uniquepgftag}{dhrystone}
    \mbox{
      \begin{tikzpicture}
    \begin{axis}[
            General,
            xmode=log,
            log basis x=2,
            xticklabel={
                    \pgfkeys{/pgf/fpu=true}
                    \pgfmathparse{2^(\tick)}%
                    \pgfmathprintnumber[fixed relative, precision=5]{\pgfmathresult}
                    \pgfkeys{/pgf/fpu=false}
                },
            xmin=512,
            xmax=65536,
            ymin=0,
            ymax=15,
            xlabel={Buffer size (entries)},
            ylabel={Slowdown $S = \frac{t_{NQC^2}}{t_{QEMU}}$},
            xlabel shift=-4pt,
            ylabel shift=-4pt,
            mark size=1pt,
        ]
        \addplot[thick, dotted, line cap = round, mark=*, mark options={solid}, color=rwth-75] table [x=cache_size, y=slowdown_cache_cnt1_merge0] {\slowdowndata}; \uniquepgflabel{pgf:1_0}
        \addplot[thick, dotted, line cap = round, mark=*, mark options={solid}, color=grun-75] table [x=cache_size, y=slowdown_cache_cnt2_merge0] {\slowdowndata}; \uniquepgflabel{pgf:2_0}
        \addplot[thick, dotted, line cap = round, mark=*, mark options={solid}, color=bordeaux-75] table [x=cache_size, y=slowdown_cache_cnt4_merge0] {\slowdowndata}; \uniquepgflabel{pgf:4_0}
        \addplot[thick, dotted, line cap = round, mark=*, mark options={solid}, color=orange-75] table [x=cache_size, y=slowdown_cache_cnt8_merge0] {\slowdowndata}; \uniquepgflabel{pgf:8_0}
        \addplot[thick, dotted, line cap = round, mark=*, mark options={solid}, color=violett-75] table [x=cache_size, y=slowdown_cache_cnt16_merge0] {\slowdowndata}; \uniquepgflabel{pgf:16_0}

        \addplot[thick, mark=*, color=rwth-75] table [x=cache_size, y=slowdown_cache_cnt1_merge1] {\slowdowndata}; \uniquepgflabel{pgf:1_1}
        \addplot[thick, mark=*, color=grun-75] table [x=cache_size, y=slowdown_cache_cnt2_merge1] {\slowdowndata}; \uniquepgflabel{pgf:2_1}
        \addplot[thick, mark=*, color=bordeaux-75] table [x=cache_size, y=slowdown_cache_cnt4_merge1] {\slowdowndata}; \uniquepgflabel{pgf:4_1}
        \addplot[thick, mark=*, color=orange-75] table [x=cache_size, y=slowdown_cache_cnt8_merge1] {\slowdowndata}; \uniquepgflabel{pgf:8_1}
        \addplot[thick, mark=*, color=violett-75] table [x=cache_size, y=slowdown_cache_cnt16_merge1] {\slowdowndata}; \uniquepgflabel{pgf:16_1}
    \end{axis}
\end{tikzpicture}
    }
    \caption{Dhrystone.}
    \label{fig:slowdown:dhrystone}
  \end{subfigure}\hfill
  \begin{subfigure}[b]{0.45\linewidth}
    \centering
    \pgfplotstableread[col sep=comma]{plot/data/results_Stream_caches_slowdown.csv}\slowdowndata
    \renewcommand{\uniquepgftag}{stream}
    \mbox{
      \begin{tikzpicture}
    \begin{axis}[
            General,
            xmode=log,
            log basis x=2,
            xticklabel={
                    \pgfkeys{/pgf/fpu=true}
                    \pgfmathparse{2^(\tick)}%
                    \pgfmathprintnumber[fixed relative, precision=5]{\pgfmathresult}
                    \pgfkeys{/pgf/fpu=false}
                },
            xmin=512,
            xmax=65536,
            ymin=0,
            ymax=15,
            xlabel={Buffer size (entries)},
            ylabel={Slowdown $S = \frac{t_{NQC^2}}{t_{QEMU}}$},
            xlabel shift=-4pt,
            ylabel shift=-4pt,
            mark size=1pt,
        ]
        \addplot[thick, dotted, line cap = round, mark=*, mark options={solid}, color=rwth-75] table [x=cache_size, y=slowdown_cache_cnt1_merge0] {\slowdowndata}; \uniquepgflabel{pgf:1_0}
        \addplot[thick, dotted, line cap = round, mark=*, mark options={solid}, color=grun-75] table [x=cache_size, y=slowdown_cache_cnt2_merge0] {\slowdowndata}; \uniquepgflabel{pgf:2_0}
        \addplot[thick, dotted, line cap = round, mark=*, mark options={solid}, color=bordeaux-75] table [x=cache_size, y=slowdown_cache_cnt4_merge0] {\slowdowndata}; \uniquepgflabel{pgf:4_0}
        \addplot[thick, dotted, line cap = round, mark=*, mark options={solid}, color=orange-75] table [x=cache_size, y=slowdown_cache_cnt8_merge0] {\slowdowndata}; \uniquepgflabel{pgf:8_0}
        \addplot[thick, dotted, line cap = round, mark=*, mark options={solid}, color=violett-75] table [x=cache_size, y=slowdown_cache_cnt16_merge0] {\slowdowndata}; \uniquepgflabel{pgf:16_0}

        \addplot[thick, mark=*, color=rwth-75] table [x=cache_size, y=slowdown_cache_cnt1_merge1] {\slowdowndata}; \uniquepgflabel{pgf:1_1}
        \addplot[thick, mark=*, color=grun-75] table [x=cache_size, y=slowdown_cache_cnt2_merge1] {\slowdowndata}; \uniquepgflabel{pgf:2_1}
        \addplot[thick, mark=*, color=bordeaux-75] table [x=cache_size, y=slowdown_cache_cnt4_merge1] {\slowdowndata}; \uniquepgflabel{pgf:4_1}
        \addplot[thick, mark=*, color=orange-75] table [x=cache_size, y=slowdown_cache_cnt8_merge1] {\slowdowndata}; \uniquepgflabel{pgf:8_1}
        \addplot[thick, mark=*, color=violett-75] table [x=cache_size, y=slowdown_cache_cnt16_merge1] {\slowdowndata}; \uniquepgflabel{pgf:16_1}
    \end{axis}
\end{tikzpicture}
    }
    \caption{Stream.}
    \label{fig:slowdown:stream}
  \end{subfigure}
  \begin{subfigure}[b]{0.45\linewidth}
    \centering
    \pgfplotstableread[col sep=comma]{plot/data/results_Whetstone_caches_slowdown.csv}\slowdowndata
    \renewcommand{\uniquepgftag}{whetstone}
    \mbox{
      \begin{tikzpicture}
    \begin{axis}[
            General,
            xmode=log,
            log basis x=2,
            xticklabel={
                    \pgfkeys{/pgf/fpu=true}
                    \pgfmathparse{2^(\tick)}%
                    \pgfmathprintnumber[fixed relative, precision=5]{\pgfmathresult}
                    \pgfkeys{/pgf/fpu=false}
                },
            xmin=512,
            xmax=65536,
            ymin=0,
            ymax=15,
            xlabel={Buffer size (entries)},
            ylabel={Slowdown $S = \frac{t_{NQC^2}}{t_{QEMU}}$},
            xlabel shift=-4pt,
            ylabel shift=-4pt,
            mark size=1pt,
        ]
        \addplot[thick, dotted, line cap = round, mark=*, mark options={solid}, color=rwth-75] table [x=cache_size, y=slowdown_cache_cnt1_merge0] {\slowdowndata}; \uniquepgflabel{pgf:1_0}
        \addplot[thick, dotted, line cap = round, mark=*, mark options={solid}, color=grun-75] table [x=cache_size, y=slowdown_cache_cnt2_merge0] {\slowdowndata}; \uniquepgflabel{pgf:2_0}
        \addplot[thick, dotted, line cap = round, mark=*, mark options={solid}, color=bordeaux-75] table [x=cache_size, y=slowdown_cache_cnt4_merge0] {\slowdowndata}; \uniquepgflabel{pgf:4_0}
        \addplot[thick, dotted, line cap = round, mark=*, mark options={solid}, color=orange-75] table [x=cache_size, y=slowdown_cache_cnt8_merge0] {\slowdowndata}; \uniquepgflabel{pgf:8_0}
        \addplot[thick, dotted, line cap = round, mark=*, mark options={solid}, color=violett-75] table [x=cache_size, y=slowdown_cache_cnt16_merge0] {\slowdowndata}; \uniquepgflabel{pgf:16_0}

        \addplot[thick, mark=*, color=rwth-75] table [x=cache_size, y=slowdown_cache_cnt1_merge1] {\slowdowndata}; \uniquepgflabel{pgf:1_1}
        \addplot[thick, mark=*, color=grun-75] table [x=cache_size, y=slowdown_cache_cnt2_merge1] {\slowdowndata}; \uniquepgflabel{pgf:2_1}
        \addplot[thick, mark=*, color=bordeaux-75] table [x=cache_size, y=slowdown_cache_cnt4_merge1] {\slowdowndata}; \uniquepgflabel{pgf:4_1}
        \addplot[thick, mark=*, color=orange-75] table [x=cache_size, y=slowdown_cache_cnt8_merge1] {\slowdowndata}; \uniquepgflabel{pgf:8_1}
        \addplot[thick, mark=*, color=violett-75] table [x=cache_size, y=slowdown_cache_cnt16_merge1] {\slowdowndata}; \uniquepgflabel{pgf:16_1}
    \end{axis}
\end{tikzpicture}
    }
    \caption{Whetstone.}
    \label{fig:slowdown:whetstone}
  \end{subfigure}
  \caption{\nqcc slowdown.}
  \label{fig:slowdown}
\end{figure}

\begin{figure*}[t]
  \centering
  \renewcommand{\uniquepgftag}{coremark}
  \begin{tikzpicture}
    \matrix [
        draw,
        fill=white,
        matrix of nodes,
        font=\footnotesize,
        anchor=south east,
        row sep=0pt,
        inner sep=0.5pt,
        column 1/.style={anchor=base east},
        line width=0.6pt,
    ] {
        \textsubscript{Merge}\textbackslash\textsuperscript{Buffers} & 1                      & 2                      & 4                      & 8                      & 16                      \\
        on                                                           & \uniquepgfref{pgf:cf1_1} & \uniquepgfref{pgf:cf2_1} & \uniquepgfref{pgf:cf4_1} & \uniquepgfref{pgf:cf8_1} & \uniquepgfref{pgf:cf16_1} \\
        off                                                          & \uniquepgfref{pgf:cf1_0} & \uniquepgfref{pgf:cf2_0} & \uniquepgfref{pgf:cf4_0} & \uniquepgfref{pgf:cf8_0} & \uniquepgfref{pgf:cf16_0} \\
    };
\end{tikzpicture}\\
  \begin{subfigure}[b]{0.24\linewidth}
    \centering
    \pgfplotstableread[col sep=comma]{plot/data/results_Coremark_caches_slowdown.csv}\slowdowndata
    \renewcommand{\uniquepgftag}{coremark}
    \mbox{
      \begin{tikzpicture}
    \def\basis{10}
    \def\mysca{5}
    \pgfplotsset{
        y coord trafo/.code={
                \pgfkeys{/pgf/fpu=true}
                \pgfmathparse{symlog(#1,\basis,\mysca)}\pgfmathresult
                \pgfkeys{/pgf/fpu=false}
            },
        y coord inv trafo/.code={
                \pgfkeys{/pgf/fpu=true}
                \pgfmathparse{symexp(#1,\basis,\mysca)}\pgfmathresult
                \pgfkeys{/pgf/fpu=false}
            },
        yticklabel style={/pgf/number format/.cd,int detect,precision=2},
    }
    \begin{axis}[
            General,
            xmode=log,
            log basis x=2,
            xticklabel={
                    \pgfkeys{/pgf/fpu=true}
                    \pgfmathparse{2^(\tick)}%
                    \pgfmathprintnumber[fixed relative, precision=5]{\pgfmathresult}
                    \pgfkeys{/pgf/fpu=false}
                },
            ytick = {0,5,10,1000,100000},
            minor ytick = {0,2.5,7.5,100,10000,1000000},
            yminorgrids=true,
            ymin=0,
            ymax=1000000,
            xmin=512,
            xmax=65536,
            xlabel={Buffer size (entries)},
            ylabel={$\#\text{All Buffers Full}$},
            xlabel shift=-4pt,
            ylabel shift=-4pt,
            mark size=1pt,
        ]
        \addplot[very thin, double, mark=none, draw=black, domain=512:65536, samples=2] {\basis};

        \addplot[thick, dotted, line cap = round, mark=*, mark options={solid}, color=rwth-75] table [x=cache_size, y=cache_full_cnt_cache_cnt1_merge0] {\slowdowndata}; \uniquepgflabel{pgf:cf1_0}
        \addplot[thick, dotted, line cap = round, mark=*, mark options={solid}, color=grun-75] table [x=cache_size, y=cache_full_cnt_cache_cnt2_merge0] {\slowdowndata}; \uniquepgflabel{pgf:cf2_0}
        \addplot[thick, dotted, line cap = round, mark=*, mark options={solid}, color=bordeaux-75] table [x=cache_size, y=cache_full_cnt_cache_cnt4_merge0] {\slowdowndata}; \uniquepgflabel{pgf:cf4_0}
        \addplot[thick, dotted, line cap = round, mark=*, mark options={solid}, color=orange-75] table [x=cache_size, y=cache_full_cnt_cache_cnt8_merge0] {\slowdowndata}; \uniquepgflabel{pgf:cf8_0}
        \addplot[thick, dotted, line cap = round, mark=*, mark options={solid}, color=violett-75] table [x=cache_size, y=cache_full_cnt_cache_cnt16_merge0] {\slowdowndata}; \uniquepgflabel{pgf:cf16_0}

        \addplot[thick, mark=*, color=rwth-75] table [x=cache_size, y=cache_full_cnt_cache_cnt1_merge1] {\slowdowndata}; \uniquepgflabel{pgf:cf1_1}
        \addplot[thick, mark=*, color=grun-75] table [x=cache_size, y=cache_full_cnt_cache_cnt2_merge1] {\slowdowndata}; \uniquepgflabel{pgf:cf2_1}
        \addplot[thick, mark=*, color=bordeaux-75] table [x=cache_size, y=cache_full_cnt_cache_cnt4_merge1] {\slowdowndata}; \uniquepgflabel{pgf:cf4_1}
        \addplot[thick, mark=*, color=orange-75] table [x=cache_size, y=cache_full_cnt_cache_cnt8_merge1] {\slowdowndata}; \uniquepgflabel{pgf:cf8_1}
        \addplot[thick, mark=*, color=violett-75] table [x=cache_size, y=cache_full_cnt_cache_cnt16_merge1] {\slowdowndata}; \uniquepgflabel{pgf:cf16_1}
    \end{axis}
\end{tikzpicture}
    }
    \caption{Coremark.}
    \label{fig:cachefull:coremark}
  \end{subfigure}\hfill
  \begin{subfigure}[b]{0.24\linewidth}
    \centering
    \pgfplotstableread[col sep=comma]{plot/data/results_Dhrystone_caches_slowdown.csv}\slowdowndata
    \renewcommand{\uniquepgftag}{dhrystone}
    \mbox{
      \begin{tikzpicture}
    \def\basis{10}
    \def\mysca{5}
    \pgfplotsset{
        y coord trafo/.code={
                \pgfkeys{/pgf/fpu=true}
                \pgfmathparse{symlog(#1,\basis,\mysca)}\pgfmathresult
                \pgfkeys{/pgf/fpu=false}
            },
        y coord inv trafo/.code={
                \pgfkeys{/pgf/fpu=true}
                \pgfmathparse{symexp(#1,\basis,\mysca)}\pgfmathresult
                \pgfkeys{/pgf/fpu=false}
            },
        yticklabel style={/pgf/number format/.cd,int detect,precision=2},
    }
    \begin{axis}[
            General,
            xmode=log,
            log basis x=2,
            xticklabel={
                    \pgfkeys{/pgf/fpu=true}
                    \pgfmathparse{2^(\tick)}%
                    \pgfmathprintnumber[fixed relative, precision=5]{\pgfmathresult}
                    \pgfkeys{/pgf/fpu=false}
                },
            ytick = {0,5,10,1000,100000},
            minor ytick = {0,2.5,7.5,100,10000,1000000},
            yminorgrids=true,
            ymin=0,
            ymax=1000000,
            xmin=512,
            xmax=65536,
            xlabel={Buffer size (entries)},
            ylabel={$\#\text{All Buffers Full}$},
            xlabel shift=-4pt,
            ylabel shift=-4pt,
            mark size=1pt,
        ]
        \addplot[very thin, double, mark=none, draw=black, domain=512:65536, samples=2] {\basis};

        \addplot[thick, dotted, line cap = round, mark=*, mark options={solid}, color=rwth-75] table [x=cache_size, y=cache_full_cnt_cache_cnt1_merge0] {\slowdowndata}; \uniquepgflabel{pgf:cf1_0}
        \addplot[thick, dotted, line cap = round, mark=*, mark options={solid}, color=grun-75] table [x=cache_size, y=cache_full_cnt_cache_cnt2_merge0] {\slowdowndata}; \uniquepgflabel{pgf:cf2_0}
        \addplot[thick, dotted, line cap = round, mark=*, mark options={solid}, color=bordeaux-75] table [x=cache_size, y=cache_full_cnt_cache_cnt4_merge0] {\slowdowndata}; \uniquepgflabel{pgf:cf4_0}
        \addplot[thick, dotted, line cap = round, mark=*, mark options={solid}, color=orange-75] table [x=cache_size, y=cache_full_cnt_cache_cnt8_merge0] {\slowdowndata}; \uniquepgflabel{pgf:cf8_0}
        \addplot[thick, dotted, line cap = round, mark=*, mark options={solid}, color=violett-75] table [x=cache_size, y=cache_full_cnt_cache_cnt16_merge0] {\slowdowndata}; \uniquepgflabel{pgf:cf16_0}

        \addplot[thick, mark=*, color=rwth-75] table [x=cache_size, y=cache_full_cnt_cache_cnt1_merge1] {\slowdowndata}; \uniquepgflabel{pgf:cf1_1}
        \addplot[thick, mark=*, color=grun-75] table [x=cache_size, y=cache_full_cnt_cache_cnt2_merge1] {\slowdowndata}; \uniquepgflabel{pgf:cf2_1}
        \addplot[thick, mark=*, color=bordeaux-75] table [x=cache_size, y=cache_full_cnt_cache_cnt4_merge1] {\slowdowndata}; \uniquepgflabel{pgf:cf4_1}
        \addplot[thick, mark=*, color=orange-75] table [x=cache_size, y=cache_full_cnt_cache_cnt8_merge1] {\slowdowndata}; \uniquepgflabel{pgf:cf8_1}
        \addplot[thick, mark=*, color=violett-75] table [x=cache_size, y=cache_full_cnt_cache_cnt16_merge1] {\slowdowndata}; \uniquepgflabel{pgf:cf16_1}
    \end{axis}
\end{tikzpicture}
    }
    \caption{Dhrystone.}
    \label{fig:cachefull:dhrystone}
  \end{subfigure}\hfill
  \begin{subfigure}[b]{0.24\linewidth}
    \centering
    \pgfplotstableread[col sep=comma]{plot/data/results_Stream_caches_slowdown.csv}\slowdowndata
    \renewcommand{\uniquepgftag}{stream}
    \mbox{
      \begin{tikzpicture}
    \def\basis{10}
    \def\mysca{5}
    \pgfplotsset{
        y coord trafo/.code={
                \pgfkeys{/pgf/fpu=true}
                \pgfmathparse{symlog(#1,\basis,\mysca)}\pgfmathresult
                \pgfkeys{/pgf/fpu=false}
            },
        y coord inv trafo/.code={
                \pgfkeys{/pgf/fpu=true}
                \pgfmathparse{symexp(#1,\basis,\mysca)}\pgfmathresult
                \pgfkeys{/pgf/fpu=false}
            },
        yticklabel style={/pgf/number format/.cd,int detect,precision=2},
    }
    \begin{axis}[
            General,
            xmode=log,
            log basis x=2,
            xticklabel={
                    \pgfkeys{/pgf/fpu=true}
                    \pgfmathparse{2^(\tick)}%
                    \pgfmathprintnumber[fixed relative, precision=5]{\pgfmathresult}
                    \pgfkeys{/pgf/fpu=false}
                },
            ytick = {0,5,10,1000,100000},
            minor ytick = {0,2.5,7.5,100,10000,1000000},
            yminorgrids=true,
            ymin=0,
            ymax=1000000,
            xmin=512,
            xmax=65536,
            xlabel={Buffer size (entries)},
            ylabel={$\#\text{All Buffers Full}$},
            xlabel shift=-4pt,
            ylabel shift=-4pt,
            mark size=1pt,
        ]
        \addplot[very thin, double, mark=none, draw=black, domain=512:65536, samples=2] {\basis};

        \addplot[thick, dotted, line cap = round, mark=*, mark options={solid}, color=rwth-75] table [x=cache_size, y=cache_full_cnt_cache_cnt1_merge0] {\slowdowndata}; \uniquepgflabel{pgf:cf1_0}
        \addplot[thick, dotted, line cap = round, mark=*, mark options={solid}, color=grun-75] table [x=cache_size, y=cache_full_cnt_cache_cnt2_merge0] {\slowdowndata}; \uniquepgflabel{pgf:cf2_0}
        \addplot[thick, dotted, line cap = round, mark=*, mark options={solid}, color=bordeaux-75] table [x=cache_size, y=cache_full_cnt_cache_cnt4_merge0] {\slowdowndata}; \uniquepgflabel{pgf:cf4_0}
        \addplot[thick, dotted, line cap = round, mark=*, mark options={solid}, color=orange-75] table [x=cache_size, y=cache_full_cnt_cache_cnt8_merge0] {\slowdowndata}; \uniquepgflabel{pgf:cf8_0}
        \addplot[thick, dotted, line cap = round, mark=*, mark options={solid}, color=violett-75] table [x=cache_size, y=cache_full_cnt_cache_cnt16_merge0] {\slowdowndata}; \uniquepgflabel{pgf:cf16_0}

        \addplot[thick, mark=*, color=rwth-75] table [x=cache_size, y=cache_full_cnt_cache_cnt1_merge1] {\slowdowndata}; \uniquepgflabel{pgf:cf1_1}
        \addplot[thick, mark=*, color=grun-75] table [x=cache_size, y=cache_full_cnt_cache_cnt2_merge1] {\slowdowndata}; \uniquepgflabel{pgf:cf2_1}
        \addplot[thick, mark=*, color=bordeaux-75] table [x=cache_size, y=cache_full_cnt_cache_cnt4_merge1] {\slowdowndata}; \uniquepgflabel{pgf:cf4_1}
        \addplot[thick, mark=*, color=orange-75] table [x=cache_size, y=cache_full_cnt_cache_cnt8_merge1] {\slowdowndata}; \uniquepgflabel{pgf:cf8_1}
        \addplot[thick, mark=*, color=violett-75] table [x=cache_size, y=cache_full_cnt_cache_cnt16_merge1] {\slowdowndata}; \uniquepgflabel{pgf:cf16_1}
    \end{axis}
\end{tikzpicture}
    }
    \caption{Stream.}
    \label{fig:cachefull:stream}
  \end{subfigure}\hfill
  \begin{subfigure}[b]{0.24\linewidth}
    \centering
    \pgfplotstableread[col sep=comma]{plot/data/results_Whetstone_caches_slowdown.csv}\slowdowndata
    \renewcommand{\uniquepgftag}{whetstone}
    \mbox{
      \begin{tikzpicture}
    \def\basis{10}
    \def\mysca{5}
    \pgfplotsset{
        y coord trafo/.code={
                \pgfkeys{/pgf/fpu=true}
                \pgfmathparse{symlog(#1,\basis,\mysca)}\pgfmathresult
                \pgfkeys{/pgf/fpu=false}
            },
        y coord inv trafo/.code={
                \pgfkeys{/pgf/fpu=true}
                \pgfmathparse{symexp(#1,\basis,\mysca)}\pgfmathresult
                \pgfkeys{/pgf/fpu=false}
            },
        yticklabel style={/pgf/number format/.cd,int detect,precision=2},
    }
    \begin{axis}[
            General,
            xmode=log,
            log basis x=2,
            xticklabel={
                    \pgfkeys{/pgf/fpu=true}
                    \pgfmathparse{2^(\tick)}%
                    \pgfmathprintnumber[fixed relative, precision=5]{\pgfmathresult}
                    \pgfkeys{/pgf/fpu=false}
                },
            ytick = {0,5,10,1000,100000},
            minor ytick = {0,2.5,7.5,100,10000,1000000},
            yminorgrids=true,
            ymin=0,
            ymax=1000000,
            xmin=512,
            xmax=65536,
            xlabel={Buffer size (entries)},
            ylabel={$\#\text{All Buffers Full}$},
            xlabel shift=-4pt,
            ylabel shift=-4pt,
            mark size=1pt,
        ]
        \addplot[very thin, double, mark=none, draw=black, domain=512:65536, samples=2] {\basis};

        \addplot[thick, dotted, line cap = round, mark=*, mark options={solid}, color=rwth-75] table [x=cache_size, y=cache_full_cnt_cache_cnt1_merge0] {\slowdowndata}; \uniquepgflabel{pgf:cf1_0}
        \addplot[thick, dotted, line cap = round, mark=*, mark options={solid}, color=grun-75] table [x=cache_size, y=cache_full_cnt_cache_cnt2_merge0] {\slowdowndata}; \uniquepgflabel{pgf:cf2_0}
        \addplot[thick, dotted, line cap = round, mark=*, mark options={solid}, color=bordeaux-75] table [x=cache_size, y=cache_full_cnt_cache_cnt4_merge0] {\slowdowndata}; \uniquepgflabel{pgf:cf4_0}
        \addplot[thick, dotted, line cap = round, mark=*, mark options={solid}, color=orange-75] table [x=cache_size, y=cache_full_cnt_cache_cnt8_merge0] {\slowdowndata}; \uniquepgflabel{pgf:cf8_0}
        \addplot[thick, dotted, line cap = round, mark=*, mark options={solid}, color=violett-75] table [x=cache_size, y=cache_full_cnt_cache_cnt16_merge0] {\slowdowndata}; \uniquepgflabel{pgf:cf16_0}

        \addplot[thick, mark=*, color=rwth-75] table [x=cache_size, y=cache_full_cnt_cache_cnt1_merge1] {\slowdowndata}; \uniquepgflabel{pgf:cf1_1}
        \addplot[thick, mark=*, color=grun-75] table [x=cache_size, y=cache_full_cnt_cache_cnt2_merge1] {\slowdowndata}; \uniquepgflabel{pgf:cf2_1}
        \addplot[thick, mark=*, color=bordeaux-75] table [x=cache_size, y=cache_full_cnt_cache_cnt4_merge1] {\slowdowndata}; \uniquepgflabel{pgf:cf4_1}
        \addplot[thick, mark=*, color=orange-75] table [x=cache_size, y=cache_full_cnt_cache_cnt8_merge1] {\slowdowndata}; \uniquepgflabel{pgf:cf8_1}
        \addplot[thick, mark=*, color=violett-75] table [x=cache_size, y=cache_full_cnt_cache_cnt16_merge1] {\slowdowndata}; \uniquepgflabel{pgf:cf16_1}
    \end{axis}
\end{tikzpicture}
    }
    \caption{Whetstone.}
    \label{fig:cachefull:whetstone}
  \end{subfigure}
  \caption{\nqcc buffer congestion.}
  \label{fig:cachefull}
  \vspace{-1em}
\end{figure*}
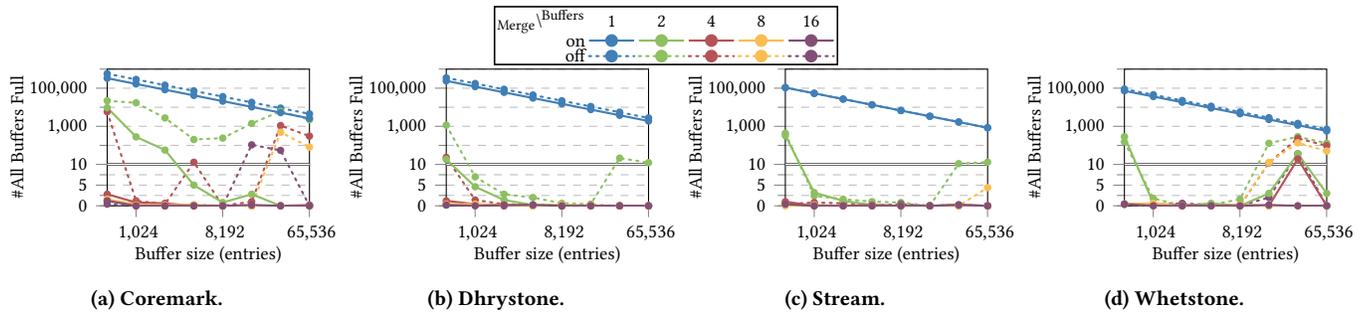

\cref{fig:slowdown} shows the ratio of the needed execution time with enabled \nqcc, $t_{NQC^2}$, to the one without \nqcc, $t_{QEMU}$, for the different benchmarks and configurations.
When only a single buffer is used (blue lines), the collector and writer cannot work in parallel which leads to a sequential behavior and thereby an extensive slowdown.
The results of the Stream and Whetstone benchmarks show that the asynchronous writer combined with multi-buffering drastically improves performance.
For those benchmarks, the gap between the blue and the other lines is the largest.
An explanation for the larger influence observed for those particular benchmarks can once again be gleaned from the data presented in \cref{fig:bench:mips}.
Workloads that reach a lower \ac{mips} number, like Stream and Whetstone, benefit the most from parallel buffer filling and flushing.
For those workloads, the execution performed by \ac{qemu} takes longer which leads to less frequent calls to \nqcc.
Less frequent calls result in more time for the writer to empty a full buffer which reduces buffer congestion.

This explanation is underlined by \cref{fig:cachefull}, which demonstrates the impact of various buffer and merge configurations on buffer congestion.
It shows how often all available buffers are full, so the collector needs to wait for the writer to empty a buffer.
Merging, as indicated in \cref{fig:bench:merge}, helps to reduce the congestion by decreasing the number of entries that need to be stored in the \ac{elog} file.

In a single-buffer setup, the collector and writer cannot work in parallel, leading to waiting every time a buffer is full.
Larger buffer sizes consistently reduce the number of filled buffers as reflected by the linear slope of the blue graphs.
Increasing the number of buffers or the buffer size typically reduces the congestion probability.
However, \cref{fig:cachefull} shows that larger buffer sizes may increase the congestion probability for some workloads.
This can be caused by the different points in time at which the buffers are emptied based on the used size.
Another reason is that larger buffers reduce the swapping overhead of the collector which lowers the amount of time the writer has to flush a buffer without congestion.
However, this slightly increased congestion probability does not lead to reduced performance as shown in \cref{fig:slowdown}.
When the congestion probability reaches zero, further increases in the buffer sizes no longer affect the performance, as a comparison between \cref{fig:slowdown,fig:cachefull} shows.

\section{Conclusion And Future Work}
\label{sec:outlook}
In the realm of software development, code coverage analysis is an essential practice that empowers developers to evaluate the effectiveness of their test suites, pinpoint untested code segments, and reveal performance bottlenecks.
We present \nqcc, a \ac{qemu}-\ac{tcg} plugin that enables instrumentation-free code coverage analysis for embedded software.
Through the use of \ac{qemu}'s plugin interface, \nqcc is also compatible with customized \ac{qemu} versions.
We presented the working principle, the structure of the \ac{elog} file format and the integration into a \ac{tcg} plugin.
Our performance optimizations, such as an asynchronous writer, the usage of multiple buffers, and the merging of blocks, can significantly reduce the slowdown.

We evaluated the performance of \nqcc using several benchmarks.
It has been seen that the slowdown of \nqcc highly depends on the executed workload.
For benchmarks that reach a high simulation speed in terms of \ac{mips}, the relative slowdown is higher.
An asynchronous writer can noticeably reduce the slowdown, especially for benchmarks that reach lower \ac{mips} values.
Depending on the \ac{tb}-execution order of the workload, the merging of entries can reduce the \ac{elog} file size and can increase the performance.

In future work, on-the-fly compression of the \ac{elog} file before writing or direct processing can be added to \nqcc to reduce the \ac{elog} file size.
In summary, \nqcc presents a versatile solution for code coverage analysis in diverse \ac{qemu} implementations, with improved performance and a deeper understanding of the factors influencing its efficiency.

\bibliographystyle{ACM-Reference-Format}
\bibliography{ref.bib}


\begin{thebibliography}{21}


\ifx \showCODEN    \undefined \def \showCODEN     #1{\unskip}     \fi
\ifx \showDOI      \undefined \def \showDOI       #1{#1}\fi
\ifx \showISBNx    \undefined \def \showISBNx     #1{\unskip}     \fi
\ifx \showISBNxiii \undefined \def \showISBNxiii  #1{\unskip}     \fi
\ifx \showISSN     \undefined \def \showISSN      #1{\unskip}     \fi
\ifx \showLCCN     \undefined \def \showLCCN      #1{\unskip}     \fi
\ifx \shownote     \undefined \def \shownote      #1{#1}          \fi
\ifx \showarticletitle \undefined \def \showarticletitle #1{#1}   \fi
\ifx \showURL      \undefined \def \showURL       {\relax}        \fi
\providecommand\bibfield[2]{#2}
\providecommand\bibinfo[2]{#2}
\providecommand\natexlab[1]{#1}
\providecommand\showeprint[2][]{arXiv:#2}

\bibitem[Bellard(2005)]%
        {bellard_qemu_2005}
\bibfield{author}{\bibinfo{person}{Fabrice Bellard}.}
  \bibinfo{year}{2005}\natexlab{}.
\newblock \showarticletitle{{QEMU}, a fast and portable dynamic translator.}.
  In \bibinfo{booktitle}{\emph{{USENIX} annual technical conference, {FREENIX}
  {Track}}}, Vol.~\bibinfo{volume}{41}. \bibinfo{publisher}{Califor-nia, USA},
  \bibinfo{pages}{10--5555}.
\newblock


\bibitem[Ben~Khadra et~al\mbox{.}(2020)]%
        {ben_khadra_efficient_2020}
\bibfield{author}{\bibinfo{person}{M.~Ammar Ben~Khadra},
  \bibinfo{person}{Dominik Stoffel}, {and} \bibinfo{person}{Wolfgang Kunz}.}
  \bibinfo{year}{2020}\natexlab{}.
\newblock \showarticletitle{Efficient binary-level coverage analysis}. In
  \bibinfo{booktitle}{\emph{Proceedings of the 28th {ACM} {Joint} {Meeting} on
  {European} {Software} {Engineering} {Conference} and {Symposium} on the
  {Foundations} of {Software} {Engineering}}}
  \emph{(\bibinfo{series}{{ESEC}/{FSE} 2020})}.
\newblock
\showISBNx{978-1-4503-7043-1}
\urldef\tempurl%
\url{https://doi.org/10.1145/3368089.3409694}
\showDOI{\tempurl}


\bibitem[Blasum et~al\mbox{.}(2007)]%
        {blasum_gcov_2007}
\bibfield{author}{\bibinfo{person}{Holger Blasum}, \bibinfo{person}{Frank
  Görgen}, {and} \bibinfo{person}{Jürgen Urban}.}
  \bibinfo{year}{2007}\natexlab{}.
\newblock \showarticletitle{Gcov on an embedded system}.
\newblock  (\bibinfo{year}{2007}).
\newblock


\bibitem[Bosbach et~al\mbox{.}(2023)]%
        {bosbach_entropy-based_2023}
\bibfield{author}{\bibinfo{person}{Nils Bosbach}, \bibinfo{person}{Lukas
  Jünger}, \bibinfo{person}{Rebecca Pelke}, \bibinfo{person}{Niko
  Zurstraßen}, {and} \bibinfo{person}{Rainer Leupers}.}
  \bibinfo{year}{2023}\natexlab{}.
\newblock \showarticletitle{Entropy-{Based} {Analysis} of {Benchmarks} for
  {Instruction} {Set} {Simulators}}. In \bibinfo{booktitle}{\emph{Proceedings
  of the {DroneSE} and {RAPIDO}: {System} {Engineering} for constrained
  embedded systems}} \emph{(\bibinfo{series}{{RAPIDO} '23})}.
\newblock
\showISBNx{9798400700453}
\urldef\tempurl%
\url{https://doi.org/10.1145/3579170.3579267}
\showDOI{\tempurl}


\bibitem[Cota and Carloni(2019)]%
        {cota_cross-isa_2019}
\bibfield{author}{\bibinfo{person}{Emilio~G. Cota} {and}
  \bibinfo{person}{Luca~P. Carloni}.} \bibinfo{year}{2019}\natexlab{}.
\newblock \showarticletitle{Cross-{ISA} machine instrumentation using fast and
  scalable dynamic binary translation}. In
  \bibinfo{booktitle}{\emph{Proceedings of the 15th {ACM} {SIGPLAN}/{SIGOPS}
  {International} {Conference} on {Virtual} {Execution} {Environments}}}.
\newblock


\bibitem[{Free Software Foundation, Inc.}(2023)]%
        {free_software_foundation_inc_gcov_2023}
\bibfield{author}{\bibinfo{person}{{Free Software Foundation, Inc.}}}
  \bibinfo{year}{2023}\natexlab{}.
\newblock \bibinfo{title}{Gcov ({Using} the {GNU} {Compiler} {Collection}
  ({GCC}))}.
\newblock
\newblock
\urldef\tempurl%
\url{https://gcc.gnu.org/onlinedocs/gcc/Gcov.html}
\showURL{%
\tempurl}


\bibitem[Gal-On and Levy(2012)]%
        {gal-on_exploring_2012}
\bibfield{author}{\bibinfo{person}{Shay Gal-On} {and} \bibinfo{person}{Markus
  Levy}.} \bibinfo{year}{2012}\natexlab{}.
\newblock \showarticletitle{Exploring coremark a benchmark maximizing
  simplicity and efficacy}.
\newblock \bibinfo{journal}{\emph{The Embedded Microprocessor Benchmark
  Consortium}} (\bibinfo{year}{2012}).
\newblock


\bibitem[Iglesias(2023)]%
        {iglesias_edgariglqemu-etrace_2023}
\bibfield{author}{\bibinfo{person}{Edgar~E. Iglesias}.}
  \bibinfo{year}{2023}\natexlab{}.
\newblock \bibinfo{title}{edgarigl/qemu-etrace}.
\newblock
\newblock
\urldef\tempurl%
\url{https://github.com/edgarigl/qemu-etrace}
\showURL{%
\tempurl}
\newblock
\shownote{original-date: 2016-02-15T09:42:55Z}.


\bibitem[Ivanković et~al\mbox{.}(2019)]%
        {ivankovic_code_2019}
\bibfield{author}{\bibinfo{person}{Marko Ivanković}, \bibinfo{person}{Goran
  Petrović}, \bibinfo{person}{René Just}, {and} \bibinfo{person}{Gordon
  Fraser}.} \bibinfo{year}{2019}\natexlab{}.
\newblock \showarticletitle{Code coverage at {Google}}. In
  \bibinfo{booktitle}{\emph{Proceedings of the 2019 27th {ACM} {Joint}
  {Meeting} on {European} {Software} {Engineering} {Conference} and {Symposium}
  on the {Foundations} of {Software} {Engineering}}}
  \emph{(\bibinfo{series}{{ESEC}/{FSE} 2019})}.
\newblock
\showISBNx{978-1-4503-5572-8}
\urldef\tempurl%
\url{https://doi.org/10.1145/3338906.3340459}
\showDOI{\tempurl}


\bibitem[{Linux Test Project}(2023)]%
        {linux_test_project_ltp_2023}
\bibfield{author}{\bibinfo{person}{{Linux Test Project}}.}
  \bibinfo{year}{2023}\natexlab{}.
\newblock \bibinfo{title}{{LTP} {GCOV} extension ({LCOV})}.
\newblock
\newblock
\urldef\tempurl%
\url{https://github.com/linux-test-project/lcov}
\showURL{%
\tempurl}
\newblock
\shownote{original-date: 2014-05-27T14:38:58Z}.


\bibitem[{LLVM Project}(2023)]%
        {llvm_project_llvm-cov_2023}
\bibfield{author}{\bibinfo{person}{{LLVM Project}}.}
  \bibinfo{year}{2023}\natexlab{}.
\newblock \bibinfo{title}{llvm-cov - emit coverage information — {LLVM}
  18.0.0git documentation}.
\newblock
\newblock
\urldef\tempurl%
\url{https://llvm.org/docs/CommandGuide/llvm-cov.html}
\showURL{%
\tempurl}


\bibitem[McCalpin(1995)]%
        {mccalpin_memory_1995}
\bibfield{author}{\bibinfo{person}{John~D. McCalpin}.}
  \bibinfo{year}{1995}\natexlab{}.
\newblock \showarticletitle{Memory {Bandwidth} and {Machine} {Balance} in
  {Current} {High} {Performance} {Computers}}.
\newblock \bibinfo{journal}{\emph{IEEE Computer Society Technical Committee on
  Computer Architecture (TCCA) Newsletter}} (\bibinfo{date}{Dec.}
  \bibinfo{year}{1995}), \bibinfo{pages}{19--25}.
\newblock


\bibitem[Miller and Maloney(1963)]%
        {miller_systematic_1963}
\bibfield{author}{\bibinfo{person}{Joan~C. Miller} {and}
  \bibinfo{person}{Clifford~J. Maloney}.} \bibinfo{year}{1963}\natexlab{}.
\newblock \showarticletitle{Systematic mistake analysis of digital computer
  programs}.
\newblock \bibinfo{journal}{\emph{Commun. ACM}} \bibinfo{volume}{6},
  \bibinfo{number}{2} (\bibinfo{date}{Feb.} \bibinfo{year}{1963}).
\newblock
\showISSN{0001-0782}


\bibitem[Nagy and Hicks(2019)]%
        {nagy_full-speed_2019}
\bibfield{author}{\bibinfo{person}{Stefan Nagy} {and} \bibinfo{person}{Matthew
  Hicks}.} \bibinfo{year}{2019}\natexlab{}.
\newblock \showarticletitle{Full-{Speed} {Fuzzing}: {Reducing} {Fuzzing}
  {Overhead} through {Coverage}-{Guided} {Tracing}}. In
  \bibinfo{booktitle}{\emph{2019 {IEEE} {Symposium} on {Security} and {Privacy}
  ({SP})}}. \bibinfo{pages}{787--802}.
\newblock
\urldef\tempurl%
\url{https://doi.org/10.1109/SP.2019.00069}
\showDOI{\tempurl}
\newblock
\shownote{ISSN: 2375-1207}.


\bibitem[{NASA Jet Propulsion Laboratory}(2023)]%
        {nasa_jet_propulsion_laboratory_nasa-jplembedded-gcov_2023}
\bibfield{author}{\bibinfo{person}{{NASA Jet Propulsion Laboratory}}.}
  \bibinfo{year}{2023}\natexlab{}.
\newblock \bibinfo{title}{nasa-jpl/embedded-gcov}.
\newblock
\newblock
\urldef\tempurl%
\url{https://github.com/nasa-jpl/embedded-gcov}
\showURL{%
\tempurl}
\newblock
\shownote{original-date: 2022-02-02T19:25:26Z}.


\bibitem[Piwowarski et~al\mbox{.}(1993)]%
        {piwowarski_coverage_1993}
\bibfield{author}{\bibinfo{person}{P. Piwowarski}, \bibinfo{person}{M. Ohba},
  {and} \bibinfo{person}{J. Caruso}.} \bibinfo{year}{1993}\natexlab{}.
\newblock \showarticletitle{Coverage measurement experience during function
  test}. In \bibinfo{booktitle}{\emph{Proceedings of 1993 15th {International}
  {Conference} on {Software} {Engineering}}}. \bibinfo{pages}{287--301}.
\newblock
\urldef\tempurl%
\url{https://doi.org/10.1109/ICSE.1993.346035}
\showDOI{\tempurl}


\bibitem[{QEMU}(2023a)]%
        {qemu_qemu_2023}
\bibfield{author}{\bibinfo{person}{{QEMU}}.} \bibinfo{year}{2023}\natexlab{a}.
\newblock \bibinfo{title}{{QEMU}}.
\newblock
\newblock
\urldef\tempurl%
\url{https://github.com/qemu/qemu}
\showURL{%
\tempurl}


\bibitem[{QEMU}(2023b)]%
        {qemu_plugin_api}
\bibfield{author}{\bibinfo{person}{{QEMU}}.} \bibinfo{year}{2023}\natexlab{b}.
\newblock \bibinfo{title}{{QEMU} {TCG} {Plugins} — {QEMU} documentation}.
\newblock
\newblock
\urldef\tempurl%
\url{https://www.qemu.org/docs/master/devel/tcg-plugins.html#api}
\showURL{%
\tempurl}


\bibitem[Weicker(1984)]%
        {weicker_dhrystone_1984}
\bibfield{author}{\bibinfo{person}{Reinhold~P. Weicker}.}
  \bibinfo{year}{1984}\natexlab{}.
\newblock \showarticletitle{Dhrystone: a synthetic systems programming
  benchmark}.
\newblock \bibinfo{journal}{\emph{Commun. ACM}} \bibinfo{volume}{27},
  \bibinfo{number}{10} (\bibinfo{date}{Oct.} \bibinfo{year}{1984}).
\newblock
\showISSN{0001-0782}
\urldef\tempurl%
\url{https://doi.org/10.1145/358274.358283}
\showDOI{\tempurl}


\bibitem[Wichmann(1970)]%
        {wichmann_statistics_1970}
\bibfield{author}{\bibinfo{person}{Brian~A Wichmann}.}
  \bibinfo{year}{1970}\natexlab{}.
\newblock \bibinfo{booktitle}{\emph{Some statistics from {ALGOL} programs}}.
\newblock \bibinfo{publisher}{Central Computer Unit, National Physical
  Laboratory}.
\newblock


\bibitem[{Xilinx}(2023)]%
        {xilinx_xilinxs_2023}
\bibfield{author}{\bibinfo{person}{{Xilinx}}.} \bibinfo{year}{2023}\natexlab{}.
\newblock \bibinfo{title}{Xilinx's fork of {QEMU}}.
\newblock
\newblock
\urldef\tempurl%
\url{https://github.com/Xilinx/qemu}
\showURL{%
\tempurl}


\end{thebibliography}

\end{document}